# Insight into the Induction Hardening Behavior of a New 0.40% C Microalloyed Steel: Effects of Initial Microstructure and Thermal Cycles

**Vahid Javaheri**[a,1], **Satish Kolli**[a], **Bjørnar Grande**[b], **David Porter**[a]

[a]Material engineering and production technology, University of Oulu, Oulu, Finland
[b]R&D, EFD Induction a.s., Skien, Norway

**Abstract**

The induction hardening behavior of a new, hot-rolled 0.4 wt.% carbon steel with the two different starting microstructures of upper and lower bainite has been simulated using a Gleeble 3800. The effect of heating rate in the range 1 - 500 °C/s on austenite grain size distribution has been rationalized. Dilatometry together with Scanning Electron Microscopy combined with Electron Backscatter Diffraction analyses and thermodynamic simulations provide insight into the austenite formation mechanisms that operate at different heating rates. Two main mechanisms of austenite formation during re-austenitization were identified: diffusional and diffusionless (massive). At conventional (1-5 °C/s) and fast (10-50 °C/s) heating rates the austenite formation mechanism and kinetics are controlled by diffusion, whereas at ultrafast heating rates (100-500 °C/s) the formation of austenite starts by diffusion control, but is later overtaken by a massive transformation mechanism. Comprehensive thermodynamic descriptions of the influence of cementite on austenite formation are discussed. The finest austenite grain size and the highest final hardness are achieved with a lower bainite starting microstructure processed with a heating rate of 50 °C/s to an austenitization temperature of 850 °C followed by cooling at 60 °C/s.

## 1. Introduction

Due to its benefits regarding cost, the environment and the mechanical properties achievable, induction heating is widely used in the hardening of a wide variety of steels. Induction hardening is usually associated with rapid heating, which affects the critical phase transformation temperatures ($A_{c1}$ and $A_{c3}$) and the prior austenite grain size as well as the microstructure and the mechanical properties after quenching [1]. Induction hardening can lead to finer martensite packet and block sizes, which can translate into improved fracture toughness by hindering cleavage crack propagation [2].

Although rapid heating is not new in heat treatment technology, e.g. in the case-hardening steel family with a carbon content of 0.35–0.6 %C, it has not been sufficiently evaluated with respect to the effect of heating rates and initial microstructure on the austenite formation mechanism, subsequent prior austenite and final grain size and consequent hardness, especially in the case of an initial bainitic

[1] Corresponding author.
Email addresses: vahid.javaheri@oulu.fi (Vahid Javaheri), satish.Kolli@oulu.fi (Satish Kolli),
Bjornar.Grande@efd-induction.com (Bjørnar Grande), David.porter@oulu.fi (David Porter)

microstructure. Esin et al. [3] using dilatometry and in-situ synchrotron X-ray diffraction tracked the phase transformation in a 0.36C-1.3Mn-0.7Cr-0.07Mo wt.% steel during continuous heating at various heating rates. They pointed out that, initially, ferrite and cementite start to transform to austenite simultaneously and then, after cementite is completely dissolved, austenite grows solely into ferrite. However, Lolla and co-workers [4] demonstrated that the dissolution of cementite is strongly affected by the heating rate and dwell time at peak temperature. By using electron microscopy, they showed that in a low-alloy steel containing Cr, even at a high peak temperature of 1100 °C, carbides did not dissolve completely when the heating rate was very high. Moreover, Danon and co-workers [5] showed that during the re-austenitisation of a martensitic steel, the prior austenite grain structure is strongly influenced by heating rate and peak temperature, such that above a critical heating rate the austenite grain size distribution becomes a mixture of fine and coarse grains. Previous research therefore shows the complexity of microstructure control in the induction hardening process. By relating thermodynamic and kinetic calculations to microstructural changes the current study attempts to develop a mechanistic understanding of the effects of heating rate and peak temperature in the case of lower and upper bainitic starting microstructures. The composition concerned is a new medium-carbon low-alloy steel microalloyed with 0.013 wt.% Nb, which has been thermomechanically processed to either upper or lower bainite.

Induction hardening has been simulated using a Gleeble 3800 thermomechanical simulator and a range of heating rates from conventional (1 and 5 °C/s), through fast (10 and 50 °C/s) to ultra-fast (100 and 500 °C/s) have been used together with austenitisation temperatures in the range 850 - 1000 °C. The work focuses on austenite transformation mechanisms and kinetics, and resultant grain size distributions. The ultimate aim of the study is a fundamental understanding of the effects of induction hardening parameters in order to achieve the highest hardness and finest grain structure for the steel in question, which has been recently developed as a potential slurry transportation pipeline material. The pipeline steel needs to have good fracture toughness as well as high hardness for the best performance against slurry erosion [6,7]. To be able to meet such a demanding requirement for the transportation pipeline system the following would be the one of best production routes: continuous casting, hot strip rolling and coiling, slitting into skelp, cold forming and high-frequency induction welding into pipe followed by induction heat treatment. Amongst all the production steps, induction welding and final induction hardening has the predominant influence on the final properties.

## 2. Experimental Methodology
### 2.1. Material and Heat Treatments

A Gleeble 3800 thermomechanical simulator has been employed in order to simulate different heating cycles during induction hardening and also to provide dilatometric phase transformation data. Two starting materials were produced using laboratory hot-rolling and direct water quenching to quench-stop temperatures (QST) of either 550 °C or 420 °C followed by cooling in a furnace to produce essentially isothermal upper or lower bainitic microstructures. The composition of the studied steel together with the initial bulk mean hardness value and mean prior austenite grain size (PAGS) are given in Table 1 More details about the material composition and as-rolled properties can be found in earlier publications of the authors [6,8].

Table 1 Composition of studied material along with its prior austenite grain size and the hardness.

| Composition (wt.%) | | | | | | | | PAGS[1] ($\mu m$) | Hardness (HV 10) | |
| --- | --- | --- | --- | --- | --- | --- | --- | --- | --- | --- |
| C | Si | Mn | Cr | Ni | Mo | Nb | N | 29±5 | QST[2] 420 | QST[2] 550 |
| 0.40 | 0.19 | 0.24 | 0.92 | 0.02 | 0.48 | 0.013 | 0.004 | | 415±5 | 360±6 |

1 Prior Austenite Grain Size expressed as mean equivalent circle diameter on a cross-section containing the rolling and plate normal directions.

2 Quench Stop Temperature in °C.

Two kinds of thermal cycles have been applied for this study: 1) different constant heating rates to 950 °C, 2) different austenitization temperatures and cooling rates. In the first set, cylindrical specimens were heated uniformly to 950 °C at constant rates of 1, 5, 10, 50, 100 and 500 °C/s and then rapidly cooled at 50 °C/s without any isothermal soaking, as schematically illustrated in Figure 1. In the second set of tests, the effect of austenitization temperature and cooling rate was studied using bars with a length of 110 mm and a diameter of 10 mm clamped into the water-cooled copper grips of the Gleeble with a free span of 35 mm. The Gleeble was programmed to produce a heating rate of 5 °C/s to a maximum temperature of 1000 °C at the mid-length of sample. Due to the cooling of the ends, this resulted in a range of peak temperatures along the length of the bar, see Figure 2. After an isothermal soaking of 30 s, samples were subjected to cooling rates of 20, 40 and 60 °C/s. To control the thermal cycle, a thermocouple was spot welded to the mid-length of the specimen (TC1). Three more thermocouples were also welded at the distances of 7, 9 and 14 mm from the mid-length to measure the temperature profile along the bar. This set-up was more efficient than using separate specimens for many peak temperature - cooling rate combinations.

2.2. Microstructure Characterization

Metallographic investigations were carried out with field emission scanning electron microscopy (FESEM) after a standard sample preparation procedure i.e. sectioning, grinding and polishing to 1 μm diamond paste, and then etching in a solution of 2% HNO3 in ethanol (2% nital) for about 10 s at room temperature. Electron backscatter diffraction (EBSD) analyses were also performed and EBSD mappings were recorded using an accelerating voltage of 15 kV, a working distance of 15 mm, a tilt angle of 70° and a step size of 0.2 μm.

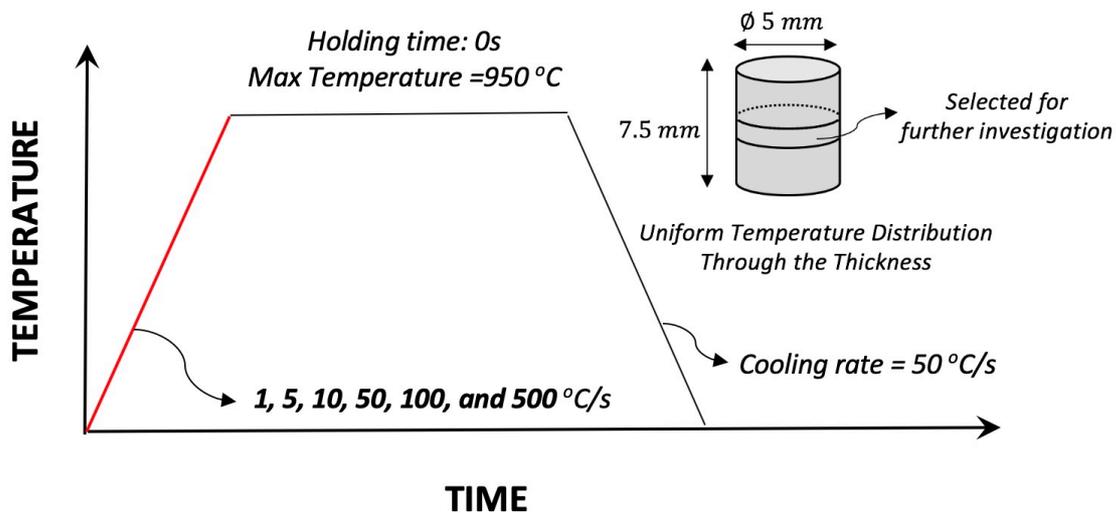

**Figure 1** Schematic illustration of thermal cycles and cylindrical sample geometry in Gleeble test set No.1.

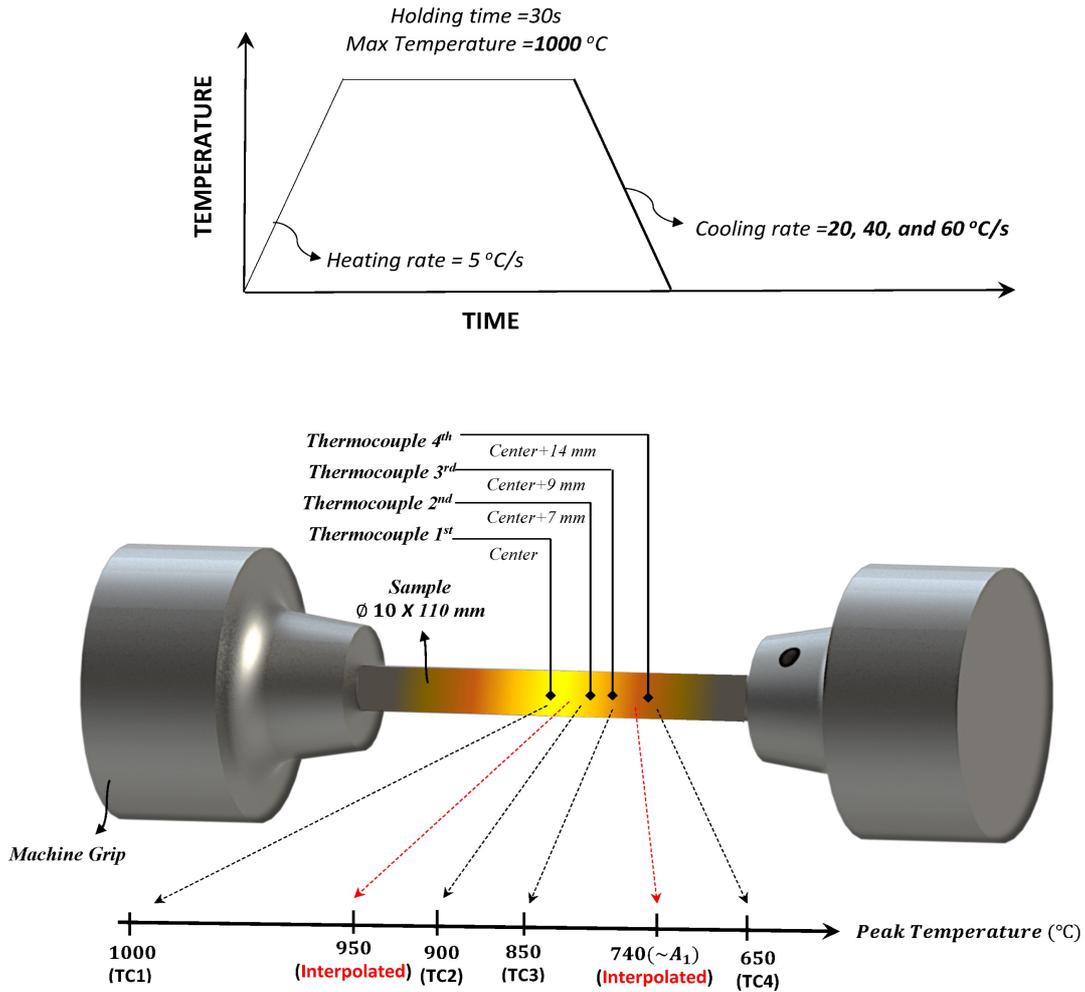

**Figure 2** Schematic illustration of Gleeble test set No.2 to apply different austenitization temperatures on one sample which has been repeated for three different cooling rates of 20, 40, and 60 °C/s.

Prior austenite grain structure was revealed with the aid of laser scanning confocal microscopy after deep etching in a saturated aqueous picric acid solution to which a few drops of detergent and hydrochloric acid had been added. To have more information about the parent austenite features, a computational reconstruction technique was also applied to the EBSD results using Matlab supplemented with the MTEX texture and crystallography analysis toolbox [9]. Briefly, the reconstruction process involved the initial assembly of grain maps from the data sets with a grain boundary tolerance of 3–5°. Subsequently, the parent austenite orientation map was constructed from this data with a two-step re-construction algorithm. Firstly, the orientation relationship between the parent austenite and product ferritic phase was determined using the method proposed by Nyyssönen et al. [9] based on the Kurdjumov–Sachs (K-S) relationship [10] (i.e., $\{111\}\gamma//\{110\}\alpha$, $<110>\gamma//<111>\alpha$). In the second step, the grain map was divided into discrete clusters using the Markov clustering method [11] proposed by Gomes and Kestens [12]. The parent austenite orientation was then calculated for each cluster separately resulting in a reconstructed austenite orientation map. The average misorientation between the reconstructed orientation for each cluster and the best fit for each individual grain was approximately 2°, indicating a good fit for the reconstructed result. An example of the reconstructed prior austenite grain structure is presented in Figure 3. Vickers hardness (HV 10) was measured on each sample after thermal cycling using a Struers Duramin A-300 hardness tester under 10 kgf load applied for 5 s and at least six measurements per sample.

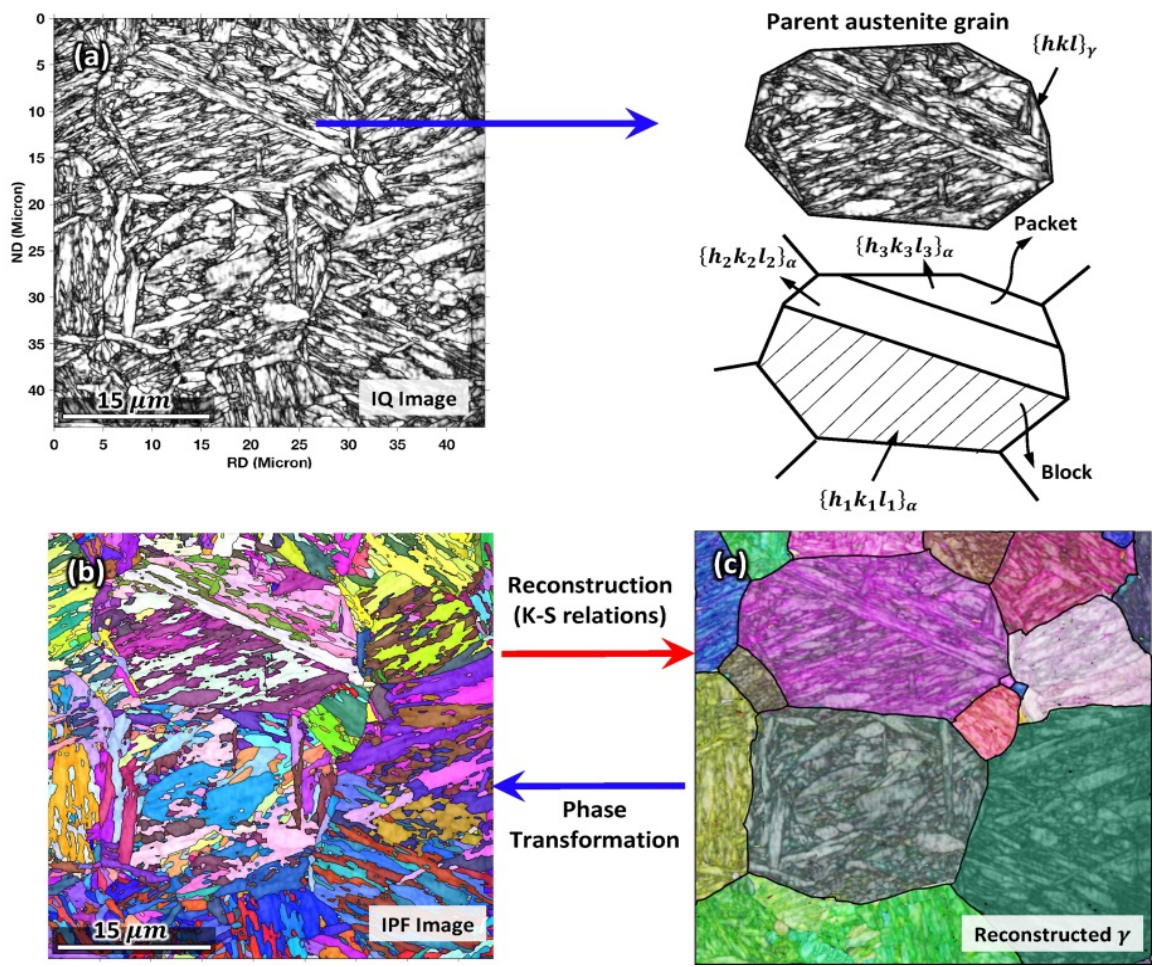

**Figure 3** An example of the reconstruction results based on a K-S orientation relation between martensite and parent austenite. a) Image quality, b) orientation map, and c) reconstructed austenite grain structure.

## 2.3. Thermodynamic and Kinetic Calculations

Thermodynamic and kinetic calculations were made using the commercial Thermo-Calc software package including the 2017b version of the DICTRA module, the TCFE9 database for thermodynamic data and the MOBFE2 database for mobility data [13]. DICTRA is a software tool for solving diffusion-controlled problems in binary and multi-components alloys involving moving phase boundaries. DICTRA calculation of the phase transformation kinetics and the speed of phase interfaces is based on a local equilibrium and diffusive flux balance at the interface [13,14].

## 3. Results and Discussion
### 3.1. Microstructures

The initial microstructures of the starting materials, i.e. upper bainite (QST550) and lower bainite (QST420) are presented in Figure 4. The microstructures consisted of a mixture of different bainitic morphologies, for instance plate-like bainite (PLB), lower lath-like bainite (LLB) and conventional upper bainite (CUB). A detailed description of the starting microstructure can be found elsewhere [8]. Comparing Figure 4(a) and Figure 4(b), it is obvious that the cementite particles in the lower bainite are finer and distributed more uniformly than in the upper bainite.

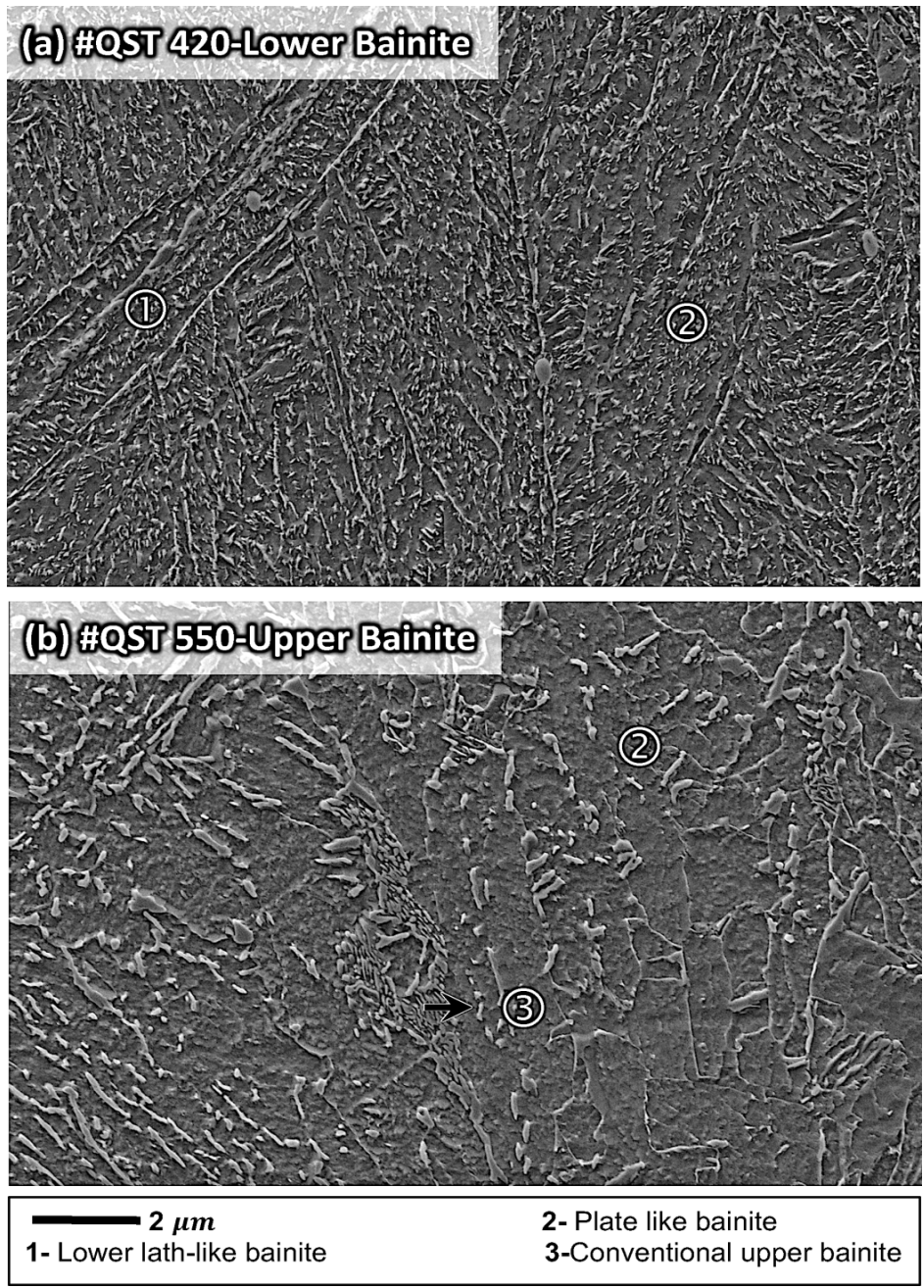

**Figure 4** Starting microstructures: a) lower bainite and b) upper bainite.

As expected, the final microstructures depended on the thermal cycle, i.e. heating rate, austenitization temperature and cooling rate. In general, the microstructures were 76 - 100 % martensite with bainite as the remaining 24 - 0 %. Examples of bainitic-martensitic and fully martensitic microstructures achieved after different treatments are given in Figure 5. More details regarding the final microstructures will be discussed later, in sections 3.2.6 and 3.4.2.

In order to gain further insight into the microstructural details, the final microstructures were also examined using EBSD. Image quality (IQ) and inverse pole figure (IPF) images clearly showed the characteristics typical of lath martensite, i.e. packets, blocks and sub-blocks, see Figure 6(a-c). It has been widely reported that during cooling, a parent austenite grain breaks down into various packets.

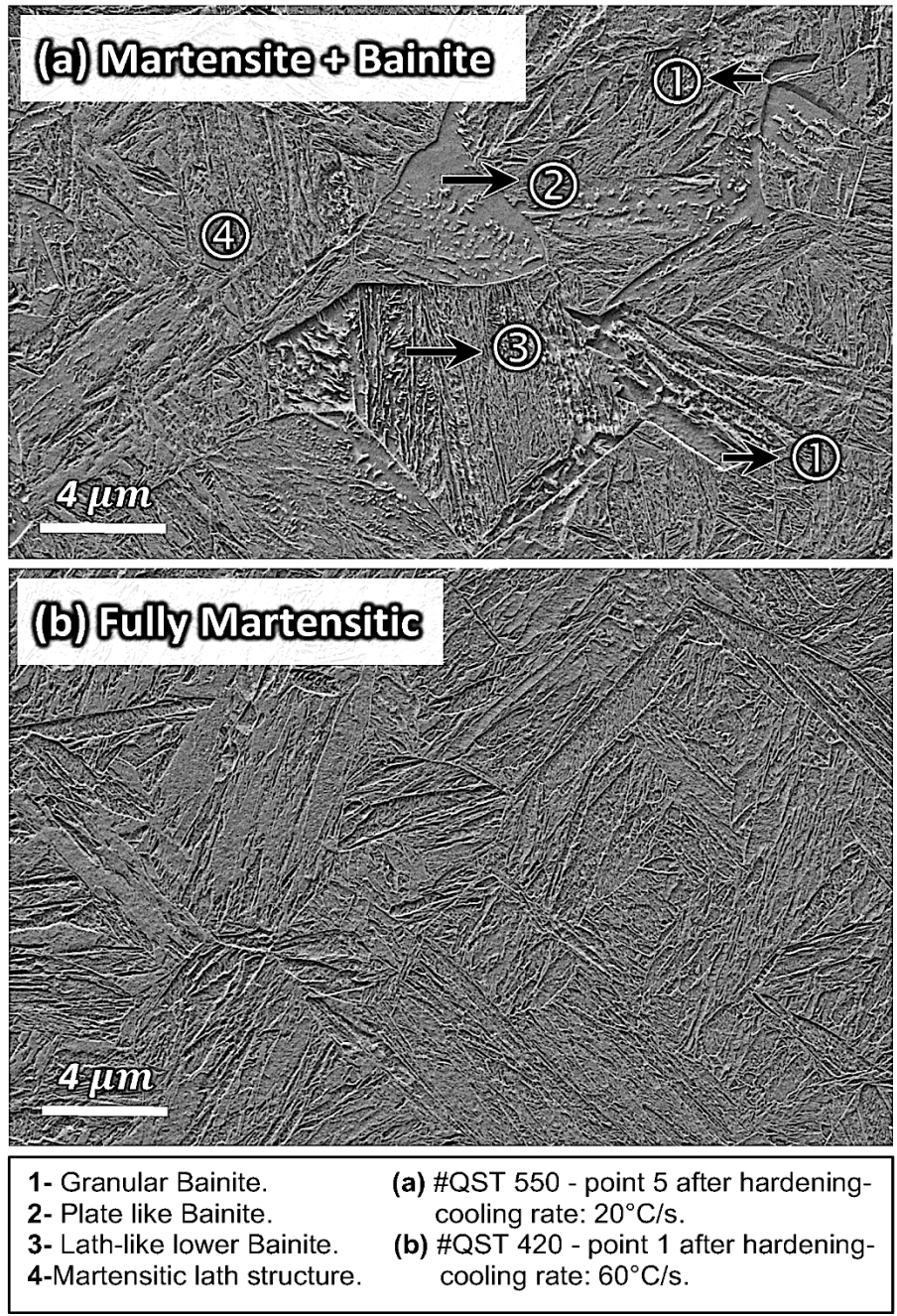

**Figure 5** Examples of as-quenched microstructures after heating at 5°C/s to 950°C and then quenching with the cooling rate of a) 20°C/s and b)60°C/s. Nital etching FESEM secondary electron image.

Each packet further subdivides to the several blocks and extended sub-blocks which contain laths. Laths, which are the finest structural units in a martensitic microstructure, are separated from each other by low-angle boundaries [15,16]. Figure 6(d-e) show the distribution of boundary misorientation angles along with a boundary map and boundary length fractions for different misorientation ranges. Three different criteria have been used to define boundaries: the black lines indicate a misorientation larger than 15°, the blue lines misorientations between 2.5° and 15°, and the red lines misorientations less than 2.5°. Following others [16,17] the lath boundaries have been characterized as having a misorientation angle less than 2.5° and the packet and block boundaries have been determined based on the misorientation more than 15°, which in fact corresponds mainly to angles in the range 47 - 60°. More details about the final microstructure grain size based on the above definition will be presented in section 3.4.2.

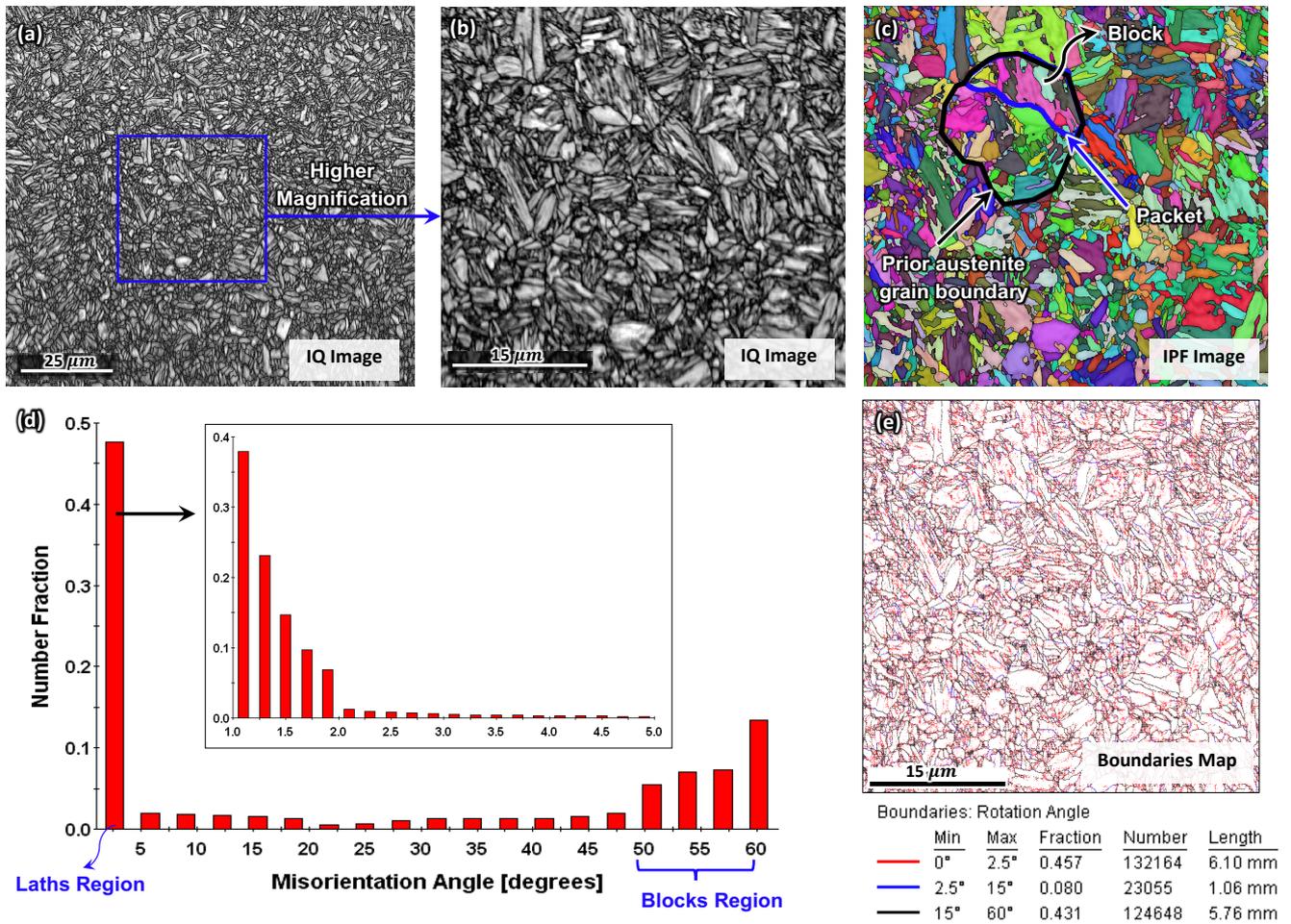

**Figure 6** a) and b) Image quality maps of an as-quenched sample (QST420) at two different magnifications. c) Inverse pole figure (IPF) orientation map (step size: 80 nm). d) Misorientation angle distribution. e) Map of boundary locations.

### 3.2. Effect of Heating Rate
#### 3.2.1. Dilatometry Results

It has been widely reported [18–20] that the start and end temperatures for austenite formation depend significantly on three main parameters, namely heating rate, initial microstructure, and chemical composition. Figure 7 shows a carbon isopleth phase diagram for the steel composition in question calculated using Thermo-Calc. According to this, on heating under equilibrium conditions for a carbon content of 0.40 wt.%, austenite formation starts (Ae1) at 742 °C and finishes (Ae3) at 788 °C.

More details concerning the predicted equilibrium phase transformation temperatures are given in Table 2. Under equilibrium conditions, cementite dissolution occurs in the temperature range between 742 and 755 °C together with an increasing volume fraction of austenite. Moreover, Nb(C,N), which is mainly NbC, is predicted to be stable up to 1102 °C. However, the initial volume fraction of Nb(C,N) is very small at 0.02 %.

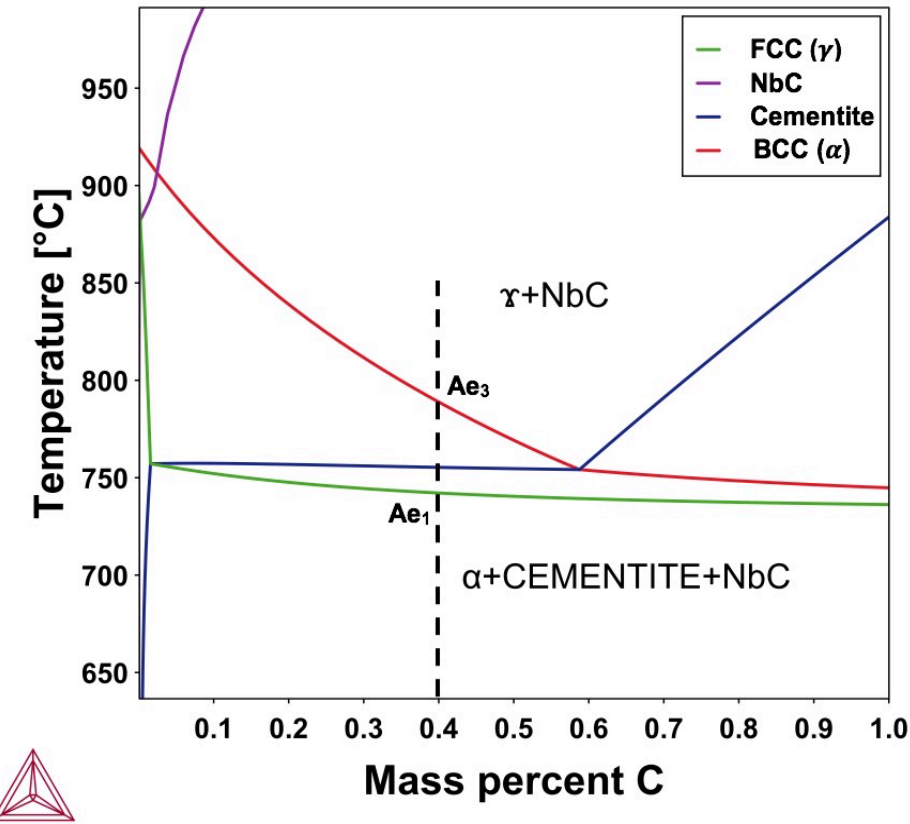

**Figure 7** Equilibrium carbon isopleth for the studied composition

Typical plots of dilation vs. temperature for different rates of heating are presented in Figure 8. The transformation temperatures $A_{c1}$ and $A_{c3}$, simply taken as the first deviation of the thermal expansion from a linear course, continuously increase with increasing the heating rate for both initial microstructures. It can be seen that for both initial bainitic microstructures all heating cycles result in 100 % austenite at the peak temperature of 950 °C. The data in Figure 8 is summarized in Figure 9, from which it can be seen that heating rate had a greater influence on the transformation temperatures of upper bainite than lower bainite and that the austenite + ferrite region spreads over a wider temperature range as the heating rate increases. Generally, $A_{c1}$ and $A_{c3}$ were higher for the material QST550. The maximum $A_{c3}$ was recorded for material QST550 at 500 °C/s: it was 886 °C, which is around 95 °C above the calculated $A_{e3}$. The differences between the measured and equilibrium transformation temperatures also indicates that during continuous heating, austenitization is noticeably slower for upper bainite than lower bainite, especially when the heating rate is slow. The reasons for this will be discussed later, in the section 3-2-3 (Cementite Dissolution).

**Table 2** Equilibrium transformation temperatures for the studied material calculated using Thermo-Calc

| Phase (volume fraction) | Temperature (°C) |
| --- | --- |
| Liquid (100%) | >1485 |
| Liquid (100%) ⇋ Liquid + γ (0.001%) | 1485 |
| Liquid + γ ⇋ γ (100%) | 1438 |
| γ (100%) ⇋ γ + Nb(C,N) (0.001%) | 1102 |
| γ + Nb(C,N) ⇋ γ (99%) + Nb(C,N) (0.016%) + α (0.001%) | 788 |
| γ + Nb(C,N) + α ⇋ γ (80%) + Nb(C,N) (0.017%) + α (19.5%) | 769 |
| γ + Nb(C,N) + α ⇋ γ (31%) + Nb(C,N) (0.018%) + α (68%) + Cementite (0.001%) | 755 |
| γ + Nb(C,N) + α + Cementite ⇋ Nb(C,N) (0.02%) + α (94%) + Cementite (5.9%) | 742 |
| Nb(C,N) (0.02%) + α (~94%) + Cementite (6.01%) | 600 |

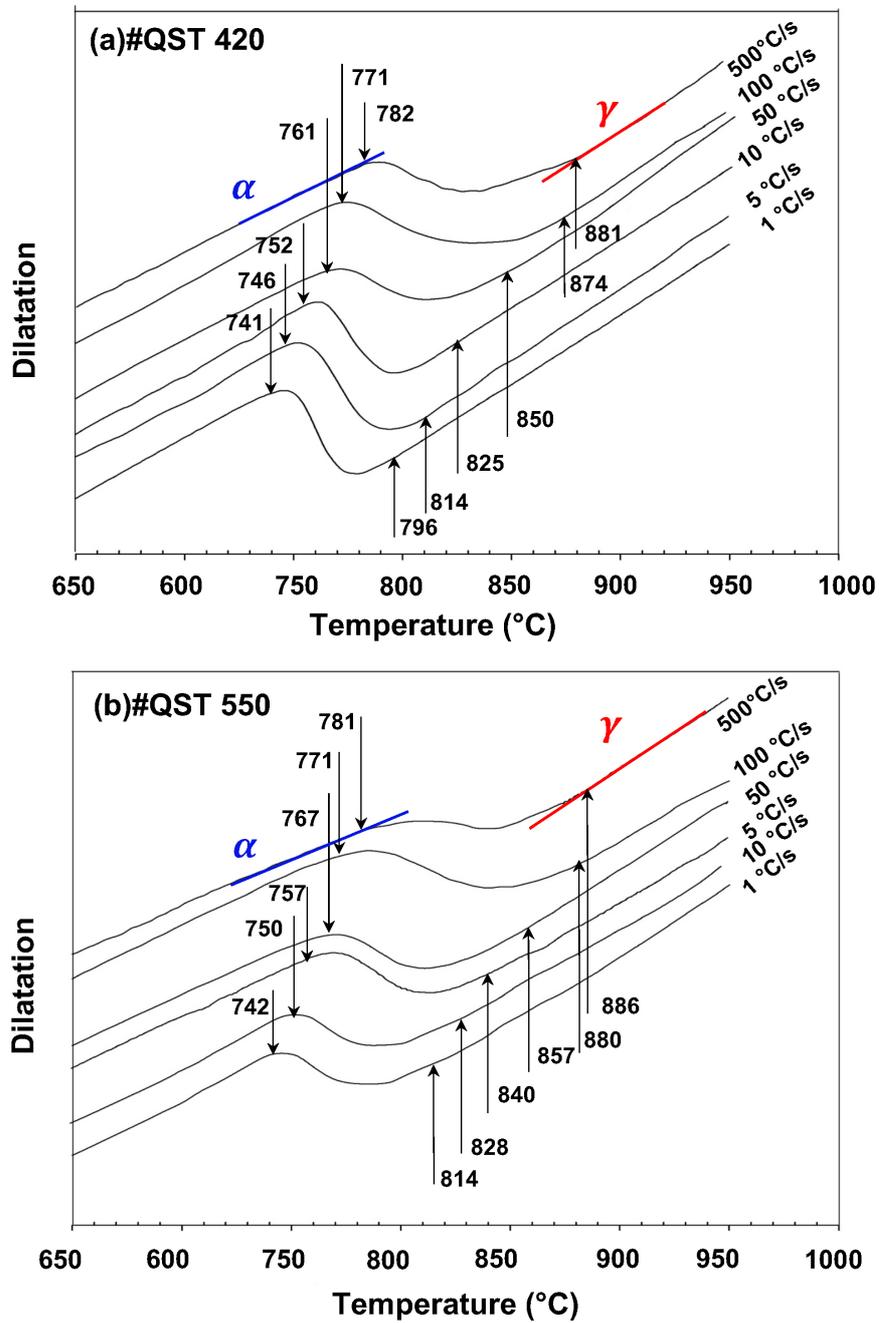

**Figure 8** Dilatometer curves obtained on heating at different rates: a) lower and b) upper bainite. Curves displaced relative to each other for clarity.

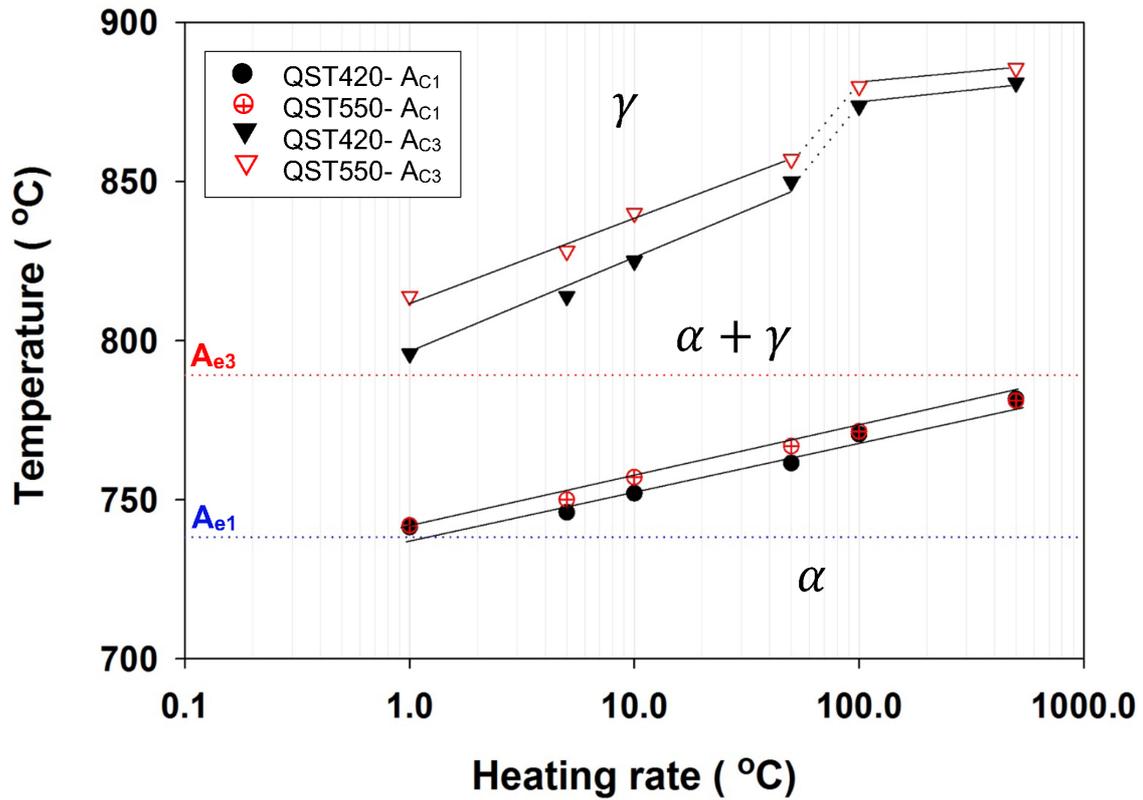

**Figure 9** Phase transformation temperatures as a function of heating rate

Figure 10 shows the estimated austenite volume fraction from the dilatometry curve using the lever rule for QST 550. It clearly demonstrates the strong sensitivity of austenite formation kinetics to increasing the heating rate in this sample. A general shift of ~30 °C can be seen between the three identical data sets obtained at conventional (slow) heating rates (1-5 °C/s), fast heating rates (10-50 °C/s), and ultra-fast heating rates (100-500 °C/s).

It is interesting to compare the observed effect of heating rate with that calculated using DICTRA. The simulation setup of austenite formation and growth during continuous heating did not converge for the multi-component steel composition due to complexity of defined system and the need to have the real composition of each phase. Therefore, a moving boundary system for a binary Fe-0.4%C was selected for the simulation. This was done by considering a spherical system comprising a central cementite particle surrounded by a very thin, initially inactive, austenite shell itself contained within a ferrite shell. The ratio of the outer diameter of the ferrite shell to the diameter of the cementite sphere was 2.55 in order to have the volume fraction of cementite required for the Fe-0.4%C system of 6 %. During heating, the austenite becomes active at $A_{e1}$. Initially the diameter of the cementite was taken as the average equivalent circle radius of the cementite particles in the QST550 specimen, i.e. 75 nm.

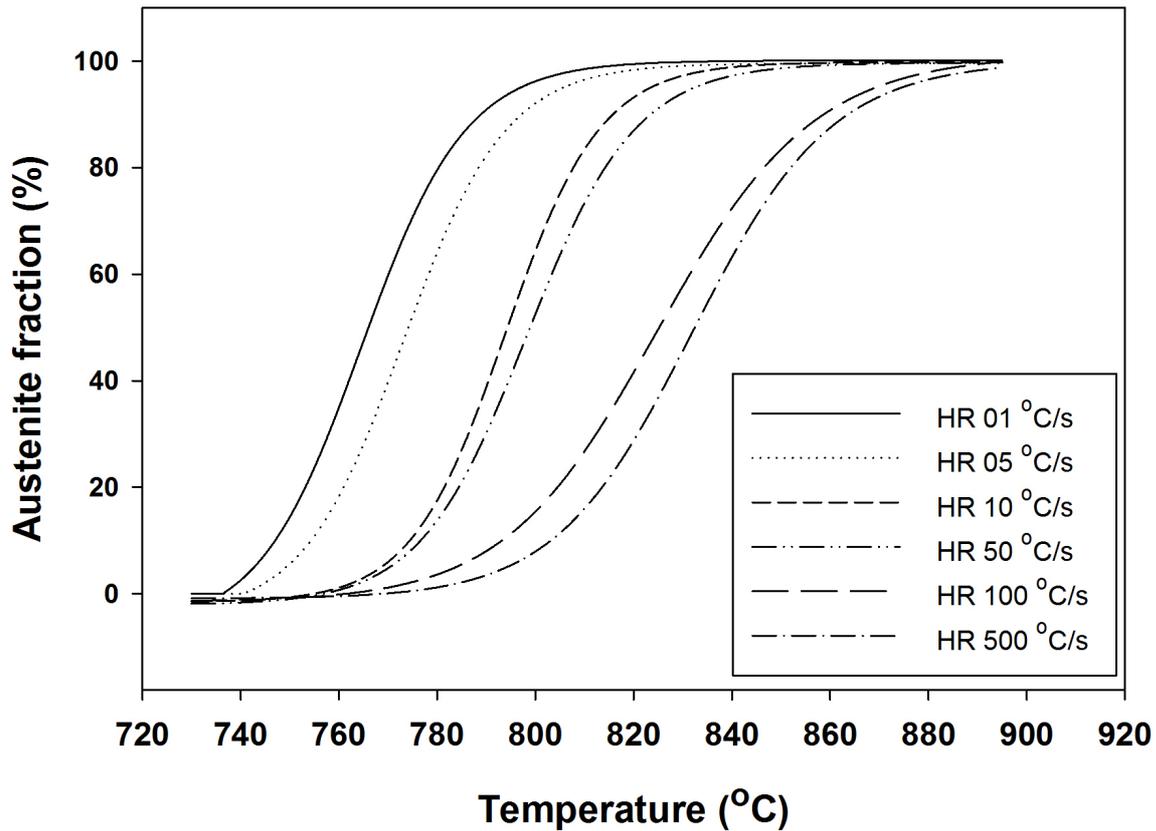

**Figure 10** The smoothed interpretation of austenite volume fraction obtained by the lever rule from the dilatometry results of sample QST 550.

Figure 11 shows the calculated mass fraction of austenite, ferrite and cementite as function of temperature during heating at different rates. The simulation shows that the formation of austenite begins with the simultaneous dissolution of cementite and ferrite that the rate at which austenite forms with increasing temperature is relatively high until the point at which the central cementite particle is completely dissolved. After this, there is a drop in the rate at which the austenite mass fraction increases with temperature. In other words, the rate of formation of austenite is high as long as cementite is present as a source of carbon for the growing austenite. It can be seen from Figure 11 that the cementite dissolution temperature and the amount of ferrite remaining at that point depends on the heating rate.

The temperature scale of the Fe-Fe3C simulations will be different from what they would be for the real alloy composition due to the missing effect of the alloying elements on the equilibrium temperatures. Also, the dissolution of cementite may be slower in the actual alloy due to the presence of Cr, which can increase the temperatures at which the cementite is fully dissolved in the growing austenite. Even so, comparing Figure 10 and Figure 11, it is clear that in the sharp change in the rate of formation of austenite predicted in Figure 11 is not observed in reality in Figure 10.

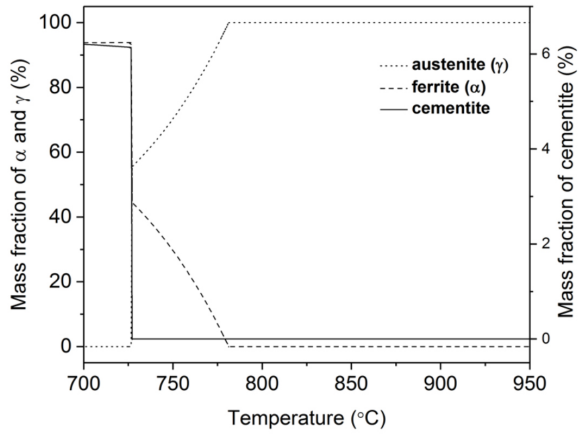
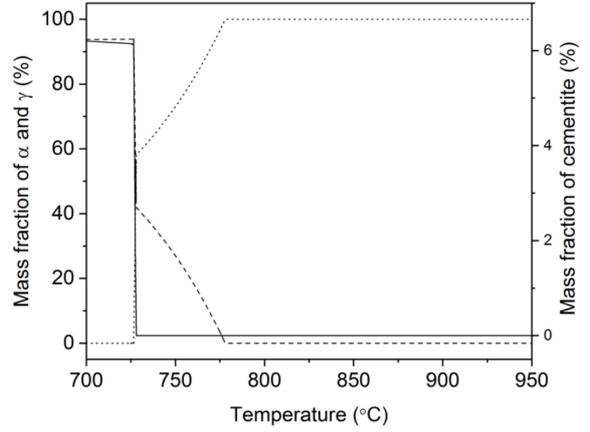
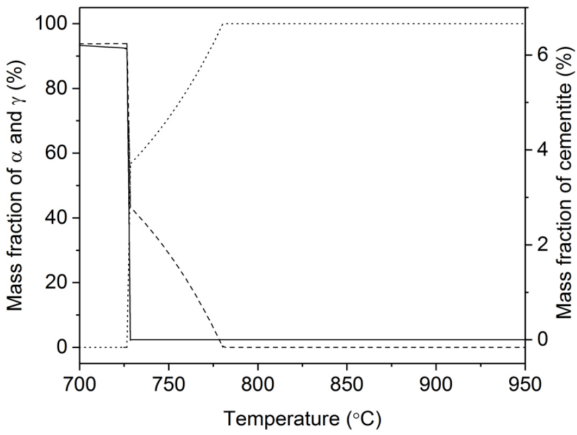
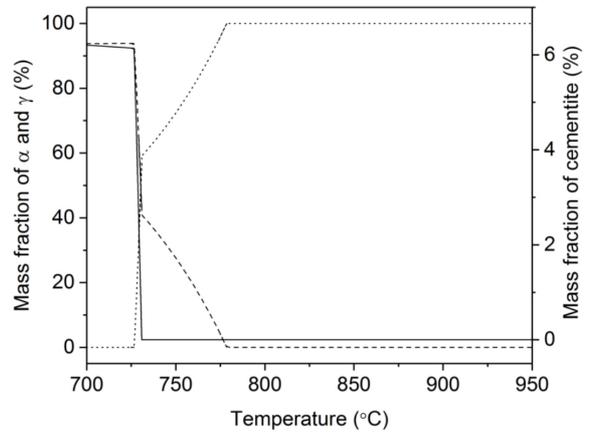
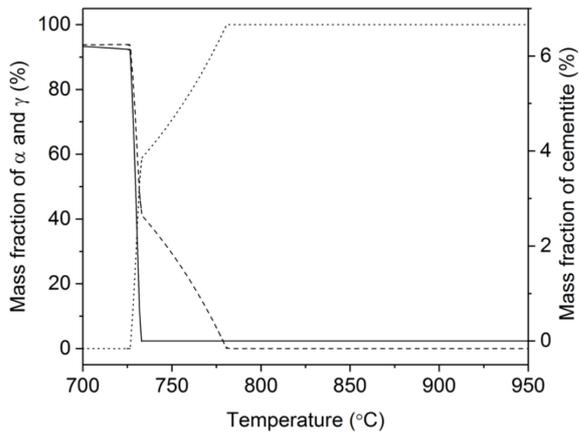
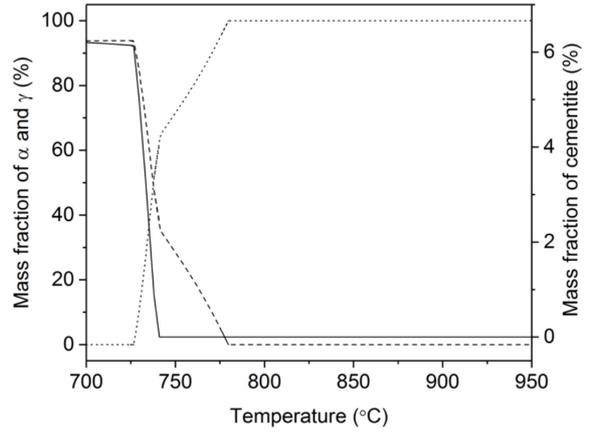

**Figure 11** DICTRA simulations of the mass fraction of austenite, ferrite and cementite during the continuous heating at heating rate of a) 1 °C/s, b) 5 °C/s, c) 10 °C/s, d) 50 °C/s, e) 100 °C/s, and f) 500 °C/s for a spherical Fe-0.4C binary system initially 192 nm in radius with a central cementite particle 75 nm in radius.

There are two main reasons for this. One is that the diameter of the cementite particles is not constant at the average, but varies from very small values up to about 300 nm. The other, more significant reason, is that the simulation assumes that austenite can form at all cementite - ferrite interfaces equally easily with zero driving force. This is not true: austenite nucleates preferentially at cementite particles associated with high-angle ferrite grain boundaries [21]. Many cementite particles may be enveloped by austenite growing from other nucleation sites. In effect, this would be roughly equivalent to an increase in the sizes of the cementite and ferrite shell in the model system. The effect of cementite diameter in the model system for a heating rate of 500°C/s is shown in Figure 12. For the small initial cementite size, the cementite dissolves in a narrow temperature window before the disappearance of ferrite, which causes a sharp two-stage austenite formation, while the dissolution of large cementite particles occurs over a much wider temperature range even extending beyond the disappearance of ferrite. Reality will be a combination of these cases leading to the observed relatively smooth variation of austenite volume fraction vs. temperature.

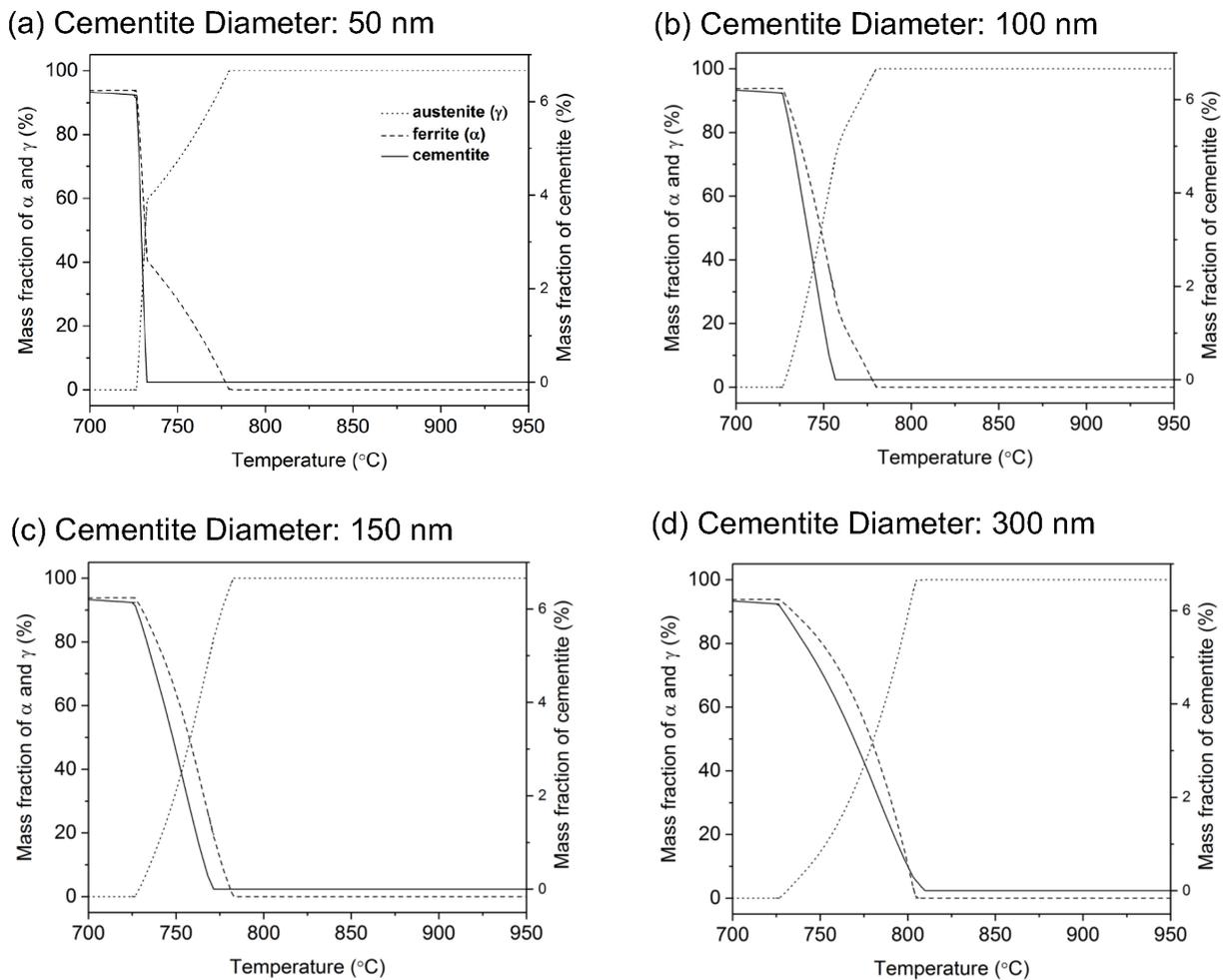

**Figure 12** Mass fraction of austenite, ferrite and cementite during the continuous heating at heating rate of 500°C/s obtained by DICTRA simulation for Fe-0.4C binary steel and initial cementite size of a)50 nm, b)100 nm, c)150 nm and d)300 nm.

### 3.2.2. Parent Austenite Grain Structure

The reconstructed prior austenite grain structures of sample QST420 heated with different rates to 950 °C and immediately cooled at 50 °C/s are shown in Figure 13. It can be seen that increasing the heating rate from to 1 to 50 °C/s leads to a refinement of the prior austenite grain size (PAGS). The average grain size, expressed as an equivalent circle diameter, for a heating rate of 50°C/s was 8.10 μm while for 1 °C/s it was 10.85 μm. However, the differences in the grain size distributions are greater than the differences in the mean values implies: the coarse grains with a diameter of 10 - 20 μm that are present after heating at 1 °C/s are absent after heating at 50 °C/s. The effect of heating rate changes, however, for rates above 50 °C/s: fine grains less than 10 μm in diameter coexist with coarse grains that are between 20 and 42 μm in diameter when the heating rate is 100 °C/s and 500 °C/s (Figure 13(f-g)). The same trend was obtained for the upper bainite material (QST550). The mean ECD prior austenite grain sizes for the both initial structures are listed in Table 3. It should be noted that after re-austenitization, the as-rolled, pancaked, starting prior austenite structure shown in Figure 13(a) is completely eliminated by all the thermal cycles irrespective of heating rate.

Prior austenite microstructures as revealed by laser confocal scanning microscopy of specimens etched with picric acid are given in Figure 14 for heating rates of 50 and 500°C/s. These show the same trend as seen with the EBSD austenite reconstruction results. Looking at the grain size distributions shown in Figure 13(b), it seems as though there are unimodal distributions for heating rates of 1 - 50 °C/s and coarse bimodal distributions for heating rates of 100 - 500 °C/s. The reasons for these differences are discussed in the following sections, 3.2.3-3.2.5.

### 3.2.3. Cementite Dissolution

During heating, austenite nucleates at cementite - ferrite interfaces and ferrite grain boundaries [22]. Particularly effective nucleation sites are formed by cementite particles on ferrite grain boundaries [21]. As the temperature increases during heating, the untransformed microstructure will experience an ever-increasing driving force for transformation, so it is likely that less effective nucleation sites will also become active such that ever more cementite is consumed in the formation and growth of new austenite grains. Cementite not consumed in this way will subsequently tend to dissolve after being enveloped by growing austenite grains. As heating rate increases, there is a tendency for cementite to survive to higher temperatures. This is apparent from Figure 11 and Figure 12 and was also reported by Clarke et al [23]. Such is the case here, with these bainitic microstructures: Figure 15 shows an example of a large undissolved cementite particle approximately 250 nm in diameter in a specimen of QST550 material heated to 950 °C at 50 °C/s and quenched.

Table 3 The influence of heating rate on the mean parent austenite grain size (ECD)

| | Heating rate (°C/s) | 1 | 5 | 10 | 50 | 100 | 500 |
|---|---|---|---|---|---|---|---|
| QST420 | PAGS[1] ($\mu m$) | 10.85 | 9.35 | 8.70 | 8.10 | 11.30 | 11.50 |
| | PAG No.[2] | 603 | 611 | 724 | 746 | 565 | 566 |
| QST550 | PAGS ($\mu m$) | 11.35 | 9.65 | 8.95 | 8.30 | 12.50 | 13.10 |
| | PAG No. | 575 | 620 | 704 | 721 | 495 | 413 |

[1] PAGS: Prior austenite grain size, [2] PAG No.: Number of prior austenite grains in an area of 220×220 $\mu m^2$

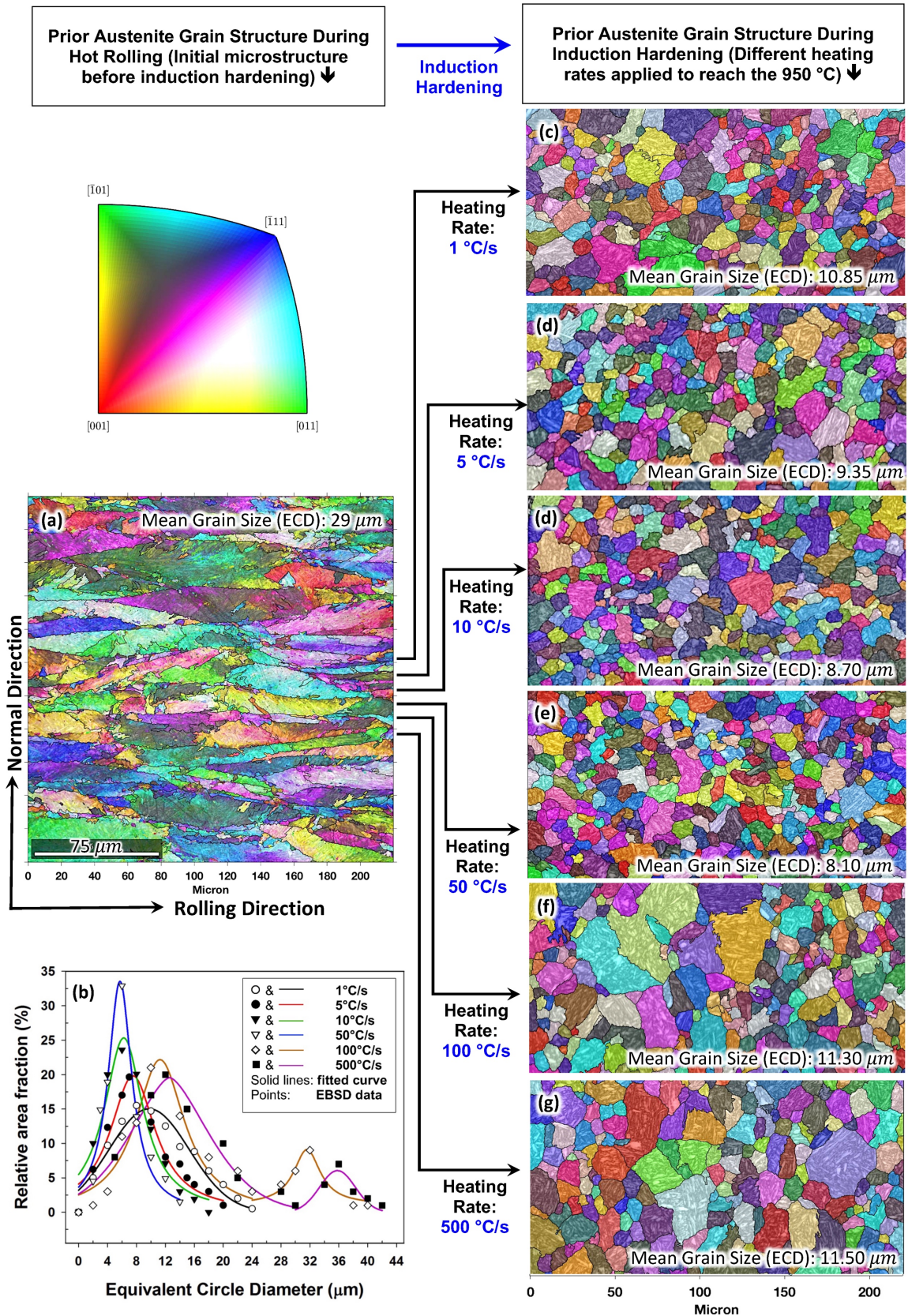

**Figure 13** a) As-rolled, pancaked parent austenite grain structure before thermal cycling. b) Grain size distribution of parent austenite after induction heating to 950°C with different heating rates followed by immediate cooling at 50 °C/s. c-g) Examples of reconstructed parent austenite grain structures given in (b). It should be noted that to fit the all structures (c-g) into this figure, they have been cropped in half from their original size of 220 $\mu m$ X 220 $\mu m$ but the quantitative data given in (b) are from the whole area.

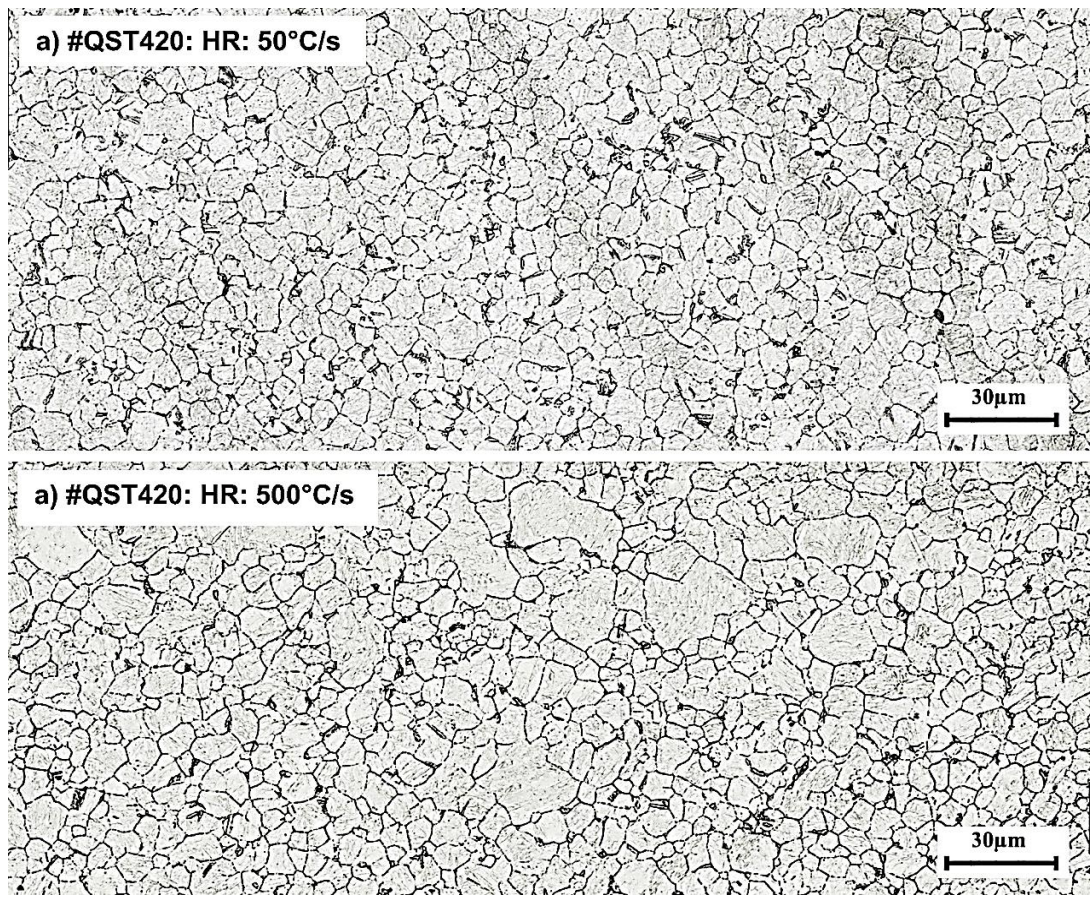

**Figure 14** Parent austenite grain structure for the sample QST420 exposed to two different heating rates as revealed by laser confocal scanning microscopy after etching with picric acid. a) 50°C/s and b) 500°C/s.

As already observed (Figure 11 and Figure 12), after heating to 950 °C, DICTRA predicts the complete dissolution of cementite particles for the binary system for all the studied heating rates. This is most likely unrealistic as it is well known that alloying elements strongly affect the phase transformation in multi-component steels. In order to simulate the effects of alloying elements on the dissolution of cementite using DICTRA it is necessary to consider a simple model system with cementite surrounded by austenite without the presence of ferrite in order to circumvent the intractable complications of a moving austenite - ferrite interface that necessitated the simple binary treatment shown in Figure 11 and Figure 12. The real condition may be somewhere in between of these two simulations (sections 3.2.1 and 3.2.3), as cementite dissolves in ferrite below $A_{c1}$ temperature and also many of cementite particles can stay surrounded by ferrite even above the $A_{c1}$ temperature. For the present simulation, cementite has been considered again as a spherical particle surrounded by the austenite matrix in a closed system. The initial cementite radius has been set to the mean experimentally measured values for each sample, i.e. radii of 50 nm and 75 nm for lower and upper bainite, respectively. The volume fractions of the phases and the initial compositions of the cementite were assumed to be those in equilibrium with ferrite at 420 or 550 °C as given by Thermo-Calc, see Table 4. With this assumption, cementite in upper bainite is predicted to contain slightly more chromium than the cementite in lower bainite, which can influence the cementite dissolution kinetics.

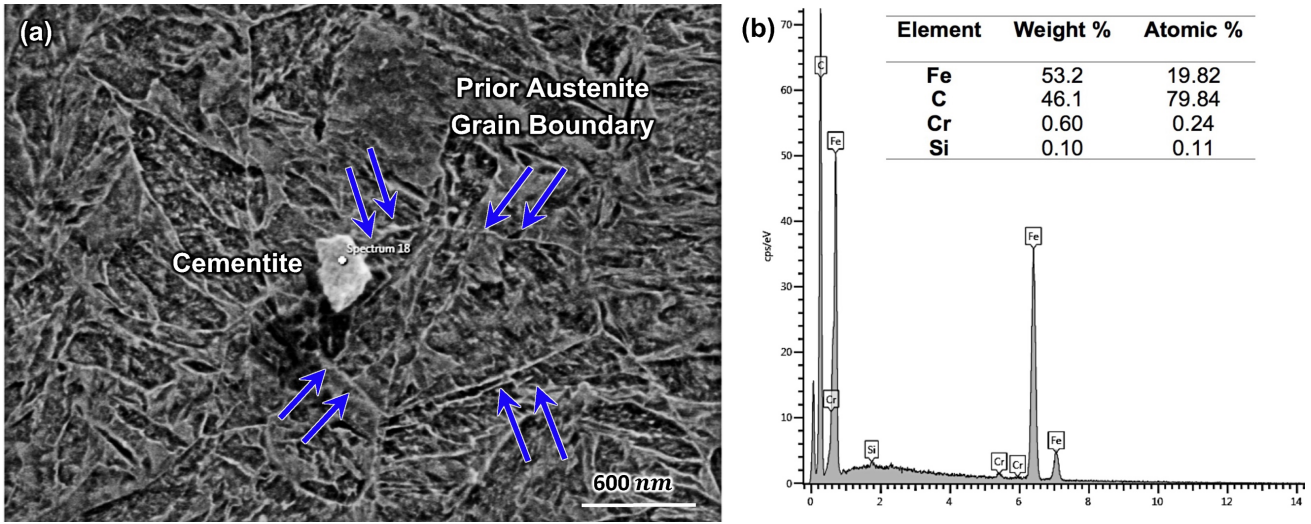

**Figure 15** Undissolved cementite particle located on a prior austenite grain boundary after re-austenitization with the heating rate of 50 °C/s and then quenching (QST550), a) FESEM image, b) EDS analysis profile of the cementite along with the details weight and atomic fraction.

Figure 16 shows the calculated cementite particle diameter during heating from room temperature to 950 °C as function of heating rate. Except for the heating rate of 1°C/s in the case of QST420 with the smallest cementite particles, the calculations predict that there should be undissolved cementite after reaching the peak temperature. The case of QST420 is also given schematically in Figure 17. It can be observed that when heating rate is really fast the cementite particle can remain undissolved at high temperature. The calculations show that cementite particles far smaller than that shown in Figure 15 and not consumed in the nucleation and early growth of austenite may survive to temperatures as high as 950 °C.

**Table 4** The considered composition for cementite in the upper and lower bainite samples which have been formed at 550°C and 420°C, respectively.

| Element | QST 420 | | QST 550 | |
| --- | --- | --- | --- | --- |
| | Mass % | Mole % | Mass % | Mole % |
| Fe | 79.21 | 63.21 | 79.87 | 63.75 |
| Cr | 8.93 | 7.65 | 9.80 | 8.40 |
| C | 6.73 | 25.00 | 6.73 | 25.00 |
| Mn | 5.07 | 4.11 | 3.38 | 2.74 |
| Mo | 0.035 | 0.016 | 0.22 | 0.10 |
| Nb | 1.91E-10 | 9.16E-11 | 9.68E-8 | 4.65E-08 |
| Si | 4.72E-13 | 7.50E-13 | 4.72E-8 | 7.50E-13 |
| Total | 100.00 | 100.00 | 100.00 | 100.00 |

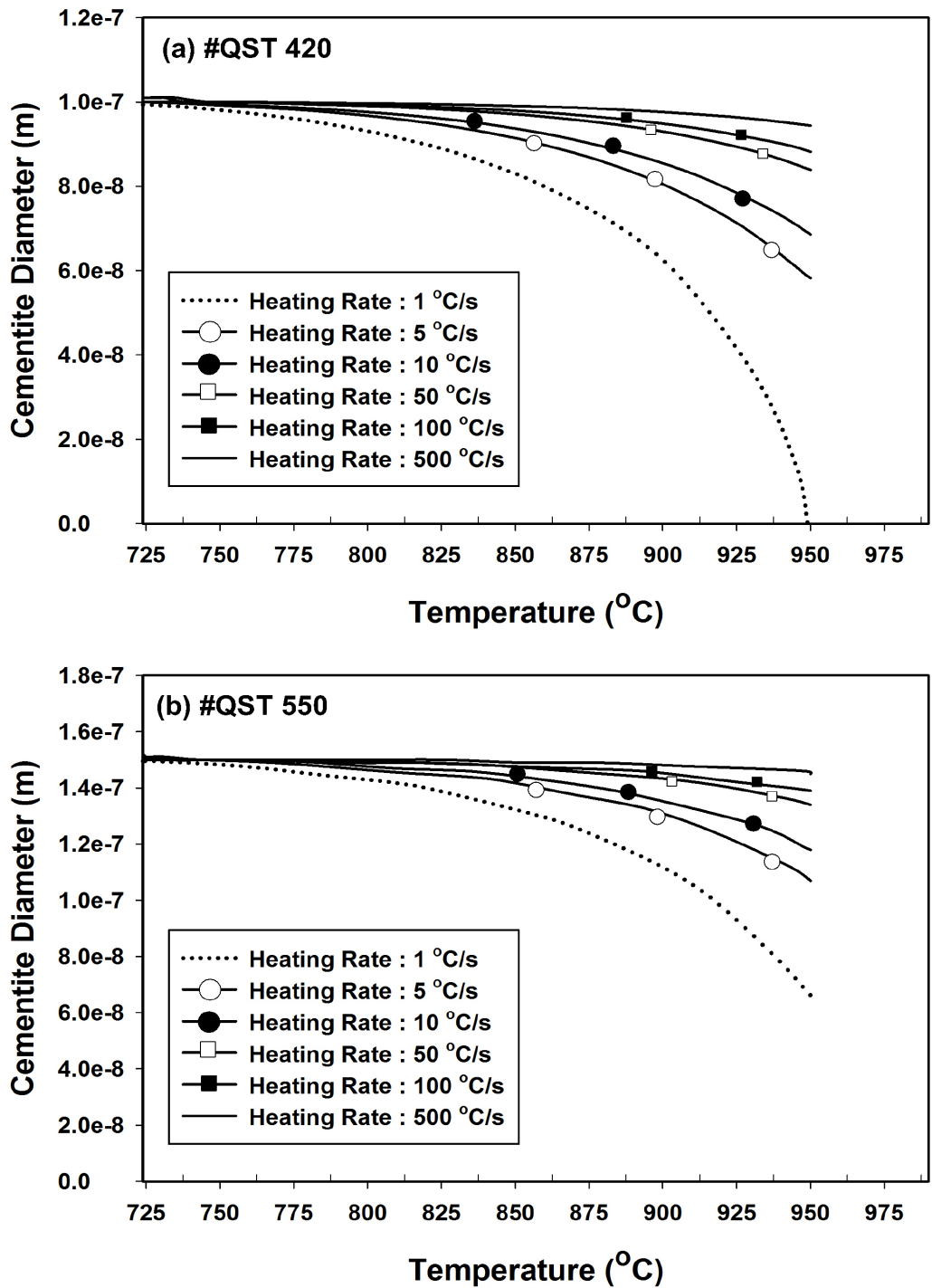

**Figure 16** The cementite dissolution in austenite as a function of temperature and heating rate to 950°C for a) lower bainite and b) upper bainite.

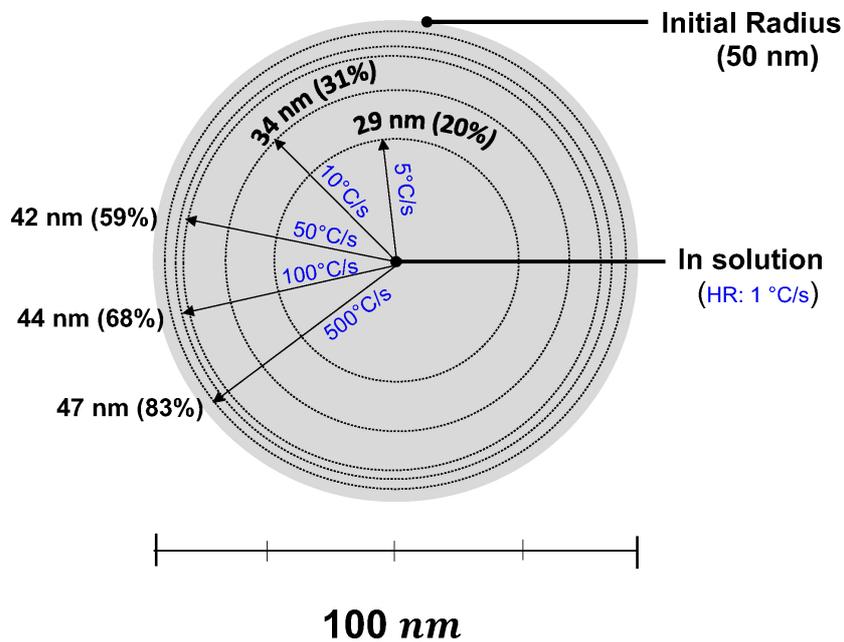

**Figure 17** Change of a single cementite particle during re-austenitization up to 950 °C with different heating rates. The percentage values in brackets indicate the volume of undissolved cementite compared to the initial volume.

Figure 18 shows that at higher heating rates cementite dissolves faster, i.e. the interface velocity is higher, but overall less cementite dissolves due to the shorter time available. The above calculations assume that the cementite has a composition that is in equilibrium with ferrite at the QST giving for example much higher Cr contents than the overall bulk content of 0.92 %. Under such conditions, dissolution is mainly controlled by the diffusion of substitutional alloying elements like Cr. If the cementite grows under paraequilibrium with no partitioning of the substitutional elements, dissolution will be much faster being controlled by interstitial C diffusion. In such a case, cementite is predicted to dissolve completely under all scenarios below the $A_{c3}$ temperatures similar to the binary condition which is controlled by the carbon diffusion. Furthermore, for all scenarios, the velocity of interface between cementite and austenite in the QST550 are slower than the QST420 samples which can be a reason for the higher $A_{c3}$ temperatures and slower austenite kinetics in the dilatometry results, especially at the lower heating rates where the diffusion plays an important role.

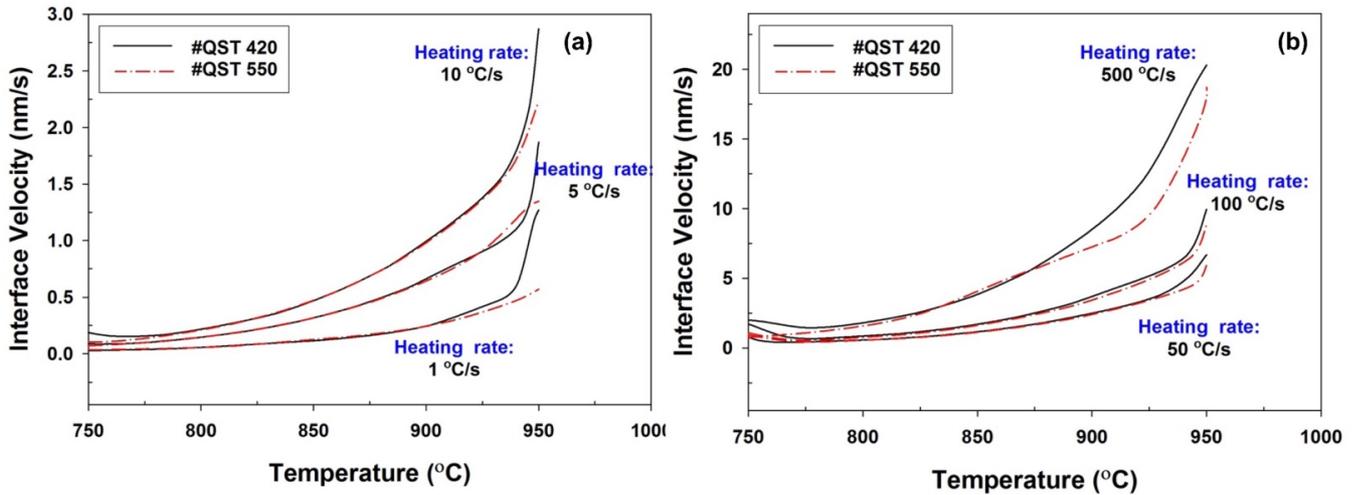

**Figure 18** The cementite - austenite interface velocity for both samples upon heating with different heating rates, a) 5 and 10°C/s and b) 50, 100 and 500°C/s. (Note different scales in a) and b))

3.2.4. Fine Unimodal Grain Size Distribution (Heating rate up to 50 °C/s)

The observed refinement of the austenite grain size with increasing heating rate up to 50 °C/s seems logical as higher heating rate reduces both the temperature window from $A_{c3}$ to the peak temperature, and the time spent crossing that temperature interval. Both of these effects reduce the extent of grain growth that can occur after the transformation is completed.

Also, higher heating rates give the opportunity for more numerous austenite nucleation sites to become active, which means a higher density of grains, i.e. a smaller initial grain size on reaching $A_{c3}$. Table 3 indicates that the number of austenite grains increased by around 20% as the heating rate increased from 1 °C/s to 50 °C/s. It is also possible that undissolved cementite particles will be more common as the heating rate increases. As their volume fraction remains significant, their pinning effect may also help retard grain growth.

3.2.5. Bimodal Grain Size Distribution (Heating rates 100 and 500 °C/s)

The appearance of a bimodal grain size distribution at the two highest studied heating rates does not follow the logic of the austenite nucleation and growth mechanisms given above and it implies that some other transformation or grain growth mechanism must become involved at high heating rates. The possibilities are i) abnormal grain growth above $A_{c3}$, ii) growth of retained austenite, or iii) the appearance of a diffusionless transformation mechanism.

It has been proposed that the incidence of abnormal grain growth is increased by high heating rates [5,24,25], but a soaking time at the peak of a few minutes is required and the peak temperature should be sufficiently high to reduce the incidence of fine precipitates in the microstructure [4]. For example, the grain coarsening temperature (GCT) for a medium carbon low alloy steel, microalloyed with V and Ti was reported about 1250 °C after holding for 3 minutes at the given temperature [24]. Such conditions are not met in this work and it is unlikely that abnormal grain growth can explain the bimodal grain structure.

It has been also pointed out that the mechanism of austenite formation and growth in the presence of retained austenite depends on the heating rate such that at slow rates the retained austenite decomposes to ferrite and carbide below $A_{c1}$, while at high heating rates the retained austenite can survive to above $A_{c1}$ where it can start to grow leading to the reformation of the austenite grains from which the retained austenite islands were remnants [4,26,27]. In this way, a slow heating rate would only lead to the formation of new austenite grains as described above, while at higher heating rates there could be a competition between new nucleation of austenite and growth from the retained austenite giving a mixture of coarse and fine grains. To test this hypothesis the following thermal cycles have been performed: heating to 740°C (below $A_{c1}$) at 5 °C/s, holding for 10 seconds, then heating to 950° C at either 100 °C/s or 500 °C/s followed by rapid quenching. Both cycles led to heterogeneous parent austenite grain structures, i.e. mixtures of coarse and fine grains very similar to those produced by direct heating at 100 and 500 °C/s. If retained austenite were the reason for the bimodal grain size distribution, the heating at 5 °C/s to 740 °C and holding 10 s would be sufficient to allow any retained austenite to decompose since direct heating at 5 °C/s results in a unimodal grain size distribution. In that case, i.e. no retained austenite remaining, the further heating beyond 740 °C at the higher rates should also produce a unimodal distribution, but such was not observed.

More evidence against the retained austenite hypothesis is provided by the size and shape of the coarse grains, which do not resemble the size and shape of the elongated prior austenite grains in the starting material as they should if they formed from the retained austenite. Therefore, the growth of retained austenite cannot be the explanation for the bimodal austenite grain size distribution seen here.

Evidence for the formation of austenite by a diffusionless, composition invariant, mechanism at high heating rates has often been proposed [19,21,22]. Diffusionless mechanisms can either be displacive, i.e. martensitic in nature, or massive. Austenite formed by a displacive transformation cannot cross high-angle ferritic grain boundaries due to the need to maintain coherency at the interface, i.e. an orientation relationship with the parent ferrite. Therefore, austenite formed in this way would need to have a grain size less than the effective, high-angle grain size of the starting bainite. In the current case, the effective grain size distribution of QST420 and QST550 is such that the largest grains have ECD values below 34 microns, which is less than the size of the coarsest austenite grains observed, i.e. 42 microns. This would seem to rule out the possibility of austenite forming by a displacive mechanism. An incoherent austenite - ferrite interface in a massive transformation, on the other hand, is not restricted in this way. Also, mobile, incoherent, massive transformation fronts can proceed at high speeds [19] and can thereby lead to the formation of coarse grains.

For the diffusionless massive transformation to be possible, the free energy of the austenite, with the composition of the ferrite into which it is growing, must be lower than that of the ferrite. This condition is reached for temperatures and ferrite compositions to the right of the $T_0$ line in Figure 19. The temperature $T_0$ is defined as temperature at which austenite and ferrite, at the same composition, have the same Gibbs free energy. From an energy point of view, ferrite can transform to austenite without the need for carbon or alloy element diffusion at temperatures above $T_0$. There is then only a need to rearrange the substitutional elements from the atomic arrangement of the bcc lattice to that of the fcc lattice at the advancing reaction front. [21,28].

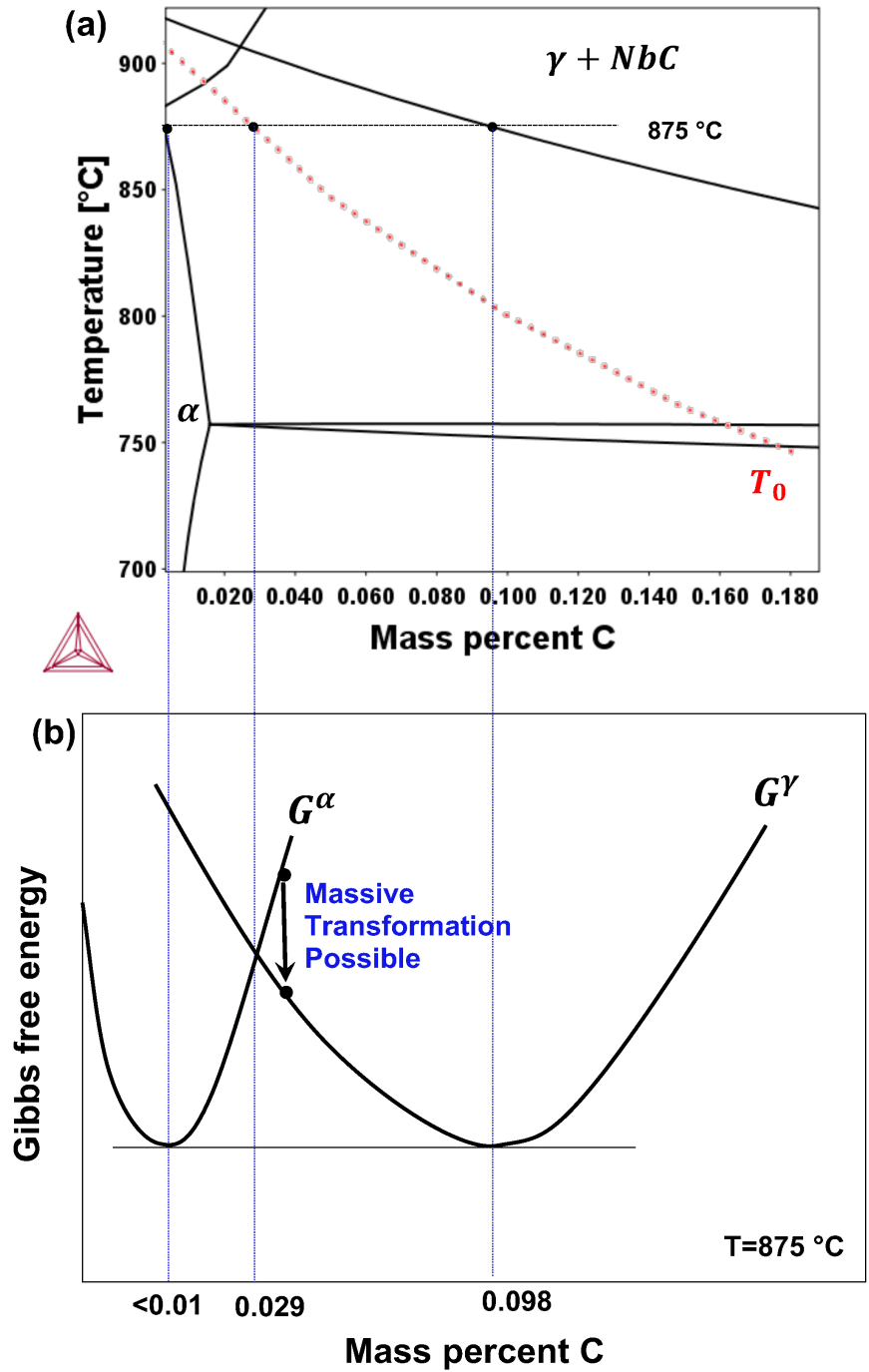

**Figure 19** a) Carbon isopleth for the studied steel composition (solid black line) together with the $T_0$ line (dotted red line). b) Gibbs free energy versus composition scheme of austenite and ferrite at temperature 875 °C.

The $A_{c3}$ temperatures in Figure 20 for the two highest heating rates suggest that the massive transformation needs to be possible at about 875 °C when the plot reaches a plateau, which means that the ferrite must contain more than $T_0$ line carbon content at this temperature which is about 0.029 wt.%. During heating, the cementite precipitates in the untransformed bainite will eventually tend to dissolve and raise the amount of carbon in solution. Assuming local equilibrium at the cementite - ferrite interface, DICTRA calculations for the dissolution of cementite during heating for the multi-component system show that even at the high rate of 500 °C/s the content of carbon in the ferrite surrounding the fine cementite precipitates (radii of 12 nm) is predicted to reach an average value above 0.029 wt.% at 875 °C, see Figure 21. It seems, therefore, as though it is feasible for the bainitic ferrite that is still untransformed at 875 °C to undergo a massive transformation to austenite. Since massive transformations can proceed behind fast-moving interfaces, this can explain the appearance of the coarse austenite grains for the two highest heating rates.

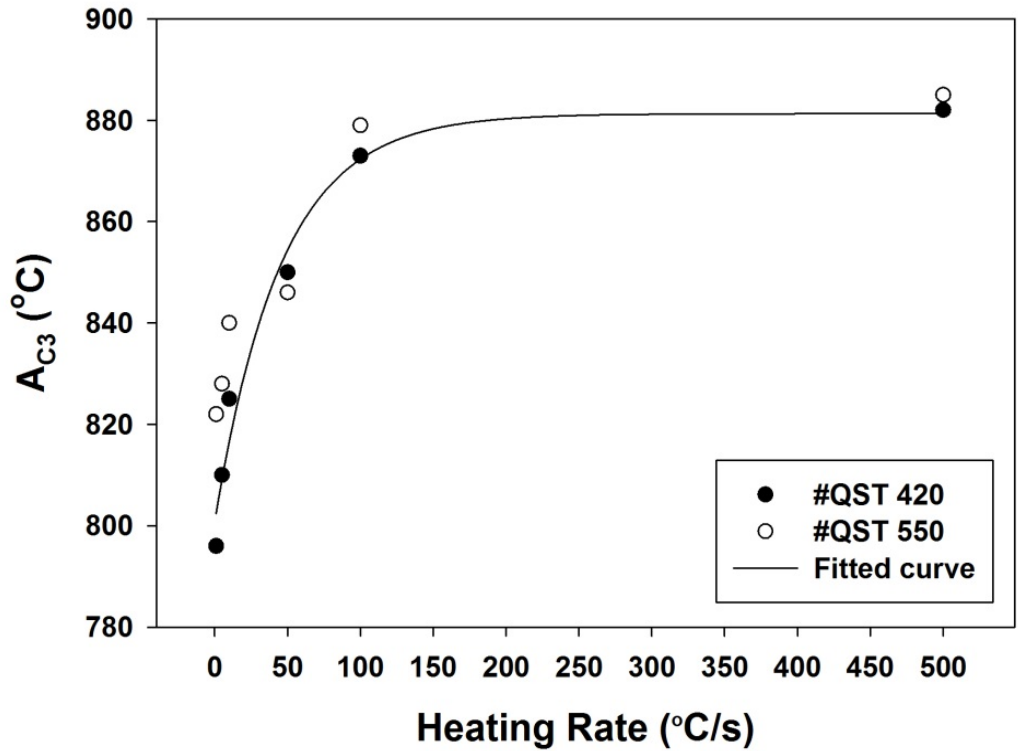

**Figure 20** $A_{c3}$ vs heating rates for both starting microstructures.

3.2.6. Effect of Heating Rate on the Final Microstructure

The martensitic microstructure of the steels after heating to 950 °C and quenching at 50 °C/s consisted of lath martensite with a high dislocation density, and laths separated primarily by low angle boundaries (as shown in Figure 6(d-e)) and arranged in a complex hierarchy of sub-blocks, blocks (a group of laths of the same orientation) and packets (a group of parallel laths with the same habit plane) within prior austenite grains [17], as given in Figure 6(a-c). The final mechanical properties of the martensitic steel including both strength and toughness is strongly affected by the lath, block and packet sizes [17,29], the finer the higher the strength and toughness. Figure 22 clearly shows that the blocks and sub-blocks in QST420 tended to be aligned in a parallel fashion with a high length to width aspect ratio, whereas the QST550 samples were characterized by more equiaxed shapes. For both samples, the variation of the widths and lengths of the packets and blocks were significant, so the equivalent circle diameter has been used to characterize block and packet sizes.

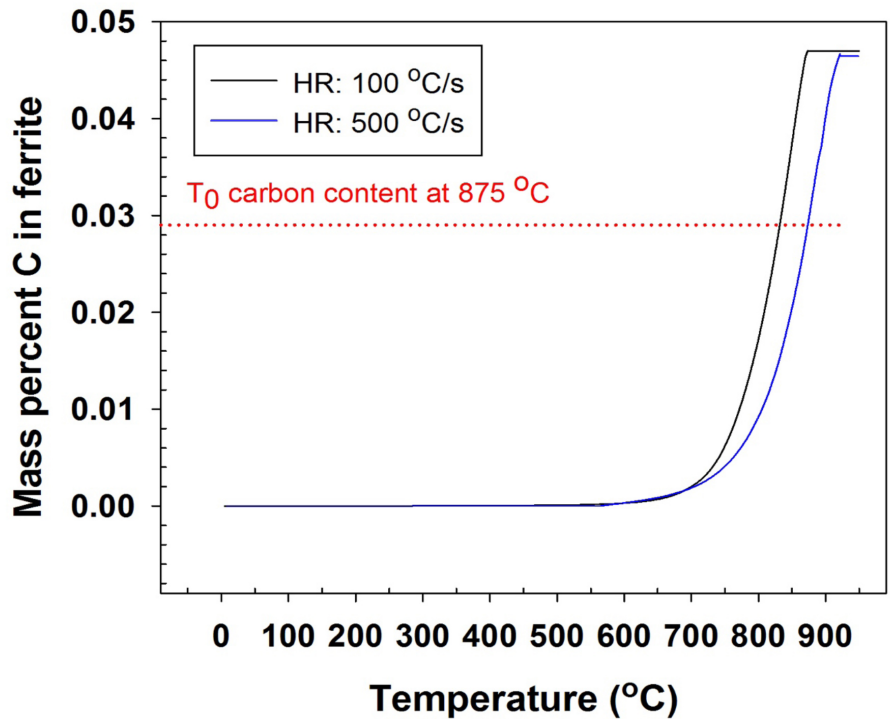

**Figure 21** The minimum carbon concentration in ferrite during the cementite dissolution obtained by DICTRA for a multi-component system and a spherical geometry consisting a center-located cementite (diameter of 24 nm) surrounded by a ferrite shell (diameter of 100 nm).

The average size of packets and blocks of the final microstructure under different heating rates and cooling rate of 50 °C/s are given in Table 5. Comparing these sizes with the parent austenite grain sizes shown in Table 3 shows that there is a correlation between the prior austenite size and the packet and block sizes such that the finest block and packet sizes were achieved for the heating rate of 50 °C/s which gave the finest parent austenite grain size in both samples. The determination of the average lath size was not completely satisfactory due to the fact that they share common boundaries with sub-blocks and blocks boundaries, but qualitatively the FESEM images revealed no obvious differences.

**Table 5** ECD sizes of packets and blocks after heating to 950 °C and cooling at 50 °C/s.

| Heating rate | QST 420 ($\mu m$) | | QST 550 ($\mu m$) | |
|---|---|---|---|---|
| | Packet | Block | Packet | Block |
| 1 | 4.23 | 2.50 | 4.17 | 2.65 |
| 5 | 4.15 | 2.60 | 4.30 | 2.55 |
| 10 | 4.40 | 2.58 | 4.32 | 2.51 |
| 50 | 3.94 | 1.84 | 3.83 | 1.95 |
| 100 | 4.86 | 2.90 | 4.80 | 3.18 |
| 500 | 4.79 | 2.85 | 4.87 | 3.07 |

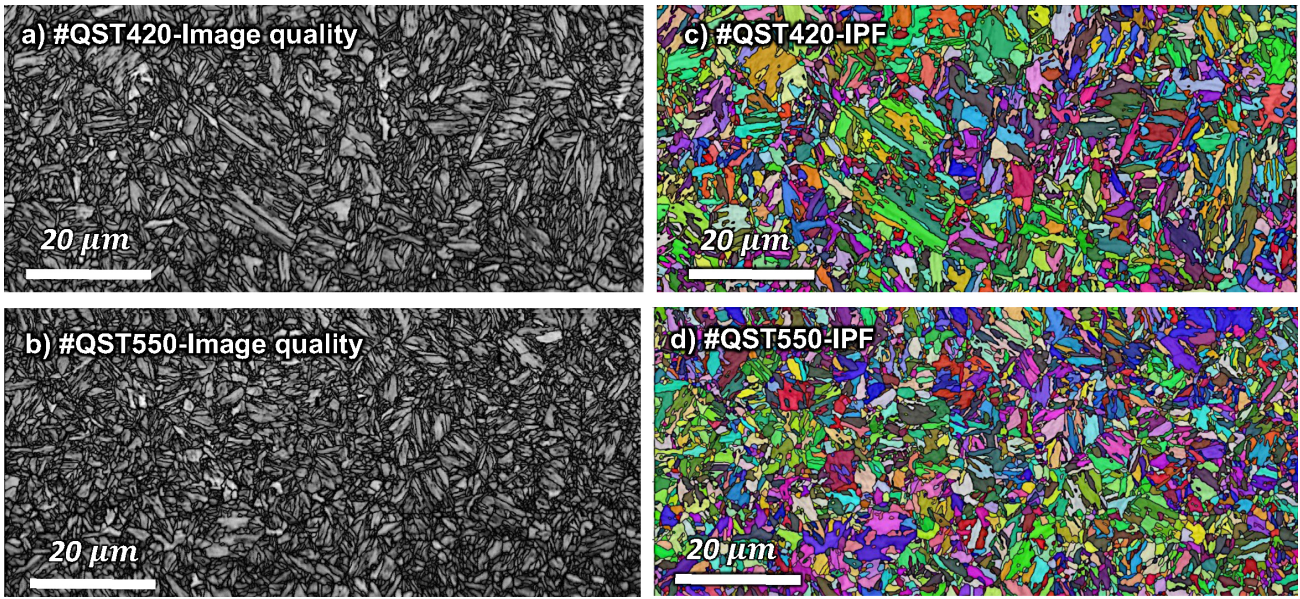

**Figure 22** a-b) Image quality and c-d) inverse pole figure (orientations map) of typical final microstructures after quenching from 950 °C showing the different martensite block morphologies. Heating rate 50°C/s.

3.3. The Effect of Austenitization Temperature

For the heating rate of 5°C/s, the effect of peak temperature on the distribution and shape of the parent austenite grains is shown in Figure 23 and Figure 24. As expected, austenite grain size increases with increasing peak temperature due to higher grain boundary mobility and more extensive grain growth. Higher temperatures will also reduce the pinning effects of any NbC precipitates or undissolved cementite leading to more grain growth at the given heating rate.

3.4. Hardness

3.4.1. Hardness versus Heating Rate

The main aim of an induction hardening process is to produce hard material. The present material is intended to achieve a hardness of about 630 HV, corresponding to 600 HB, as it has been reported that good performance can be achieved with martensite having about this level of hardness when erosion resistant pipeline is sought [30]. Mean hardness versus heating rate is presented in Figure 25. The hardness gradually increases with increasing heating rate reaching a peak of around 650HV at the heating rate of 50°C/s after which it decreases for the heating rates of 100°C/s and 500°C/s, most probably due to the coarser grain structure resulting from those heating rates. Moreover, martensite hardness is principally a function of carbon content and it is expected that the homogeneity of the carbon concentration in the austenite, and the martensite formed from it, will decrease as the heating rate increases causing increased hardness fluctuations through the sample. This was clearly observed in the experiments. For example, as given in Figure 25., the standard deviation for heating rates of 100 and 500°C/s are considerably higher than for the other heating rates. The same trend has been reported for 5150 carbon steel [23].

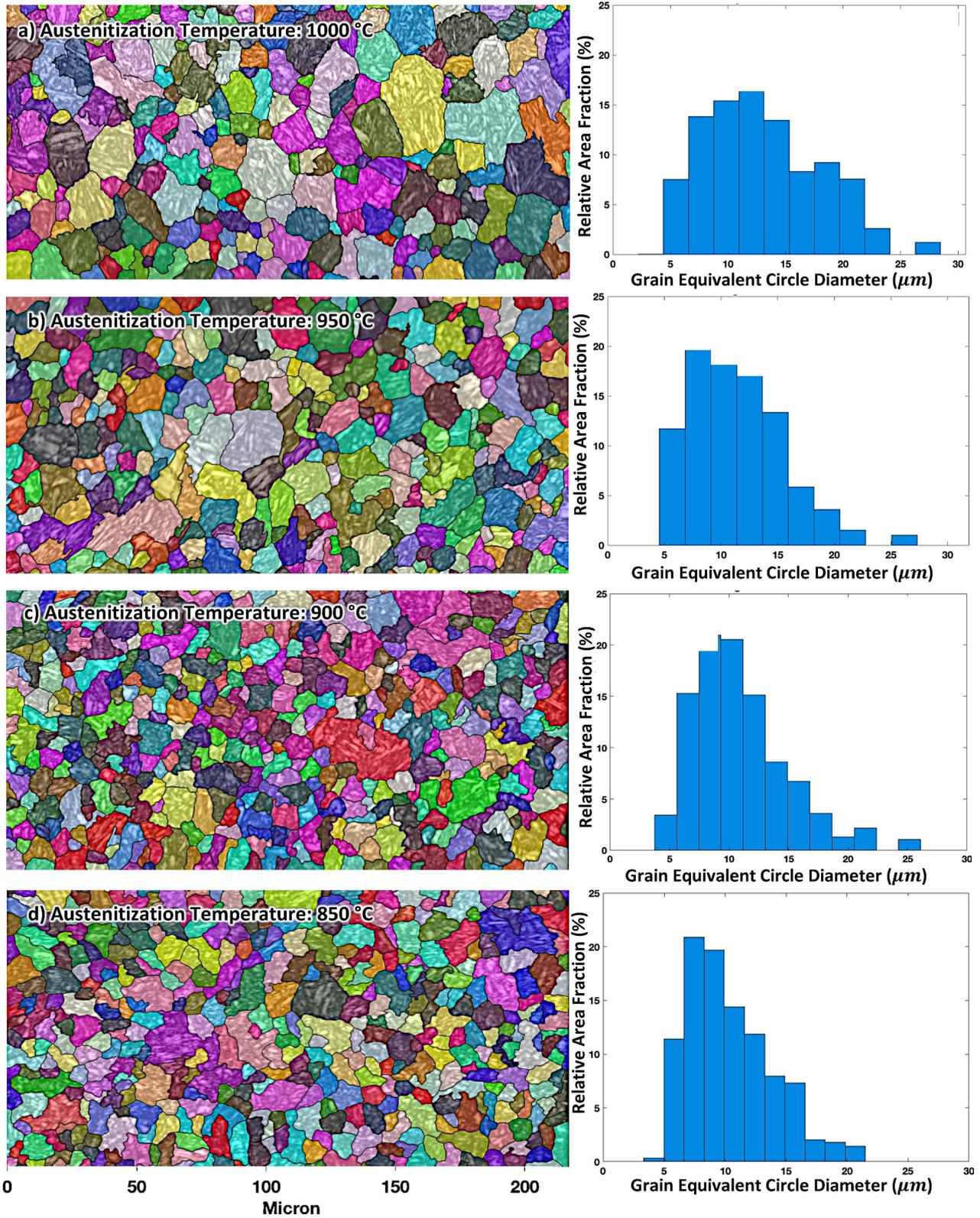

**Figure 23** The parent austenite grain structure and grain size distribution for the lower bainite sample after heating to austenitization temperatures a) 1000, b) 950, c) 900, and d) 850°C.

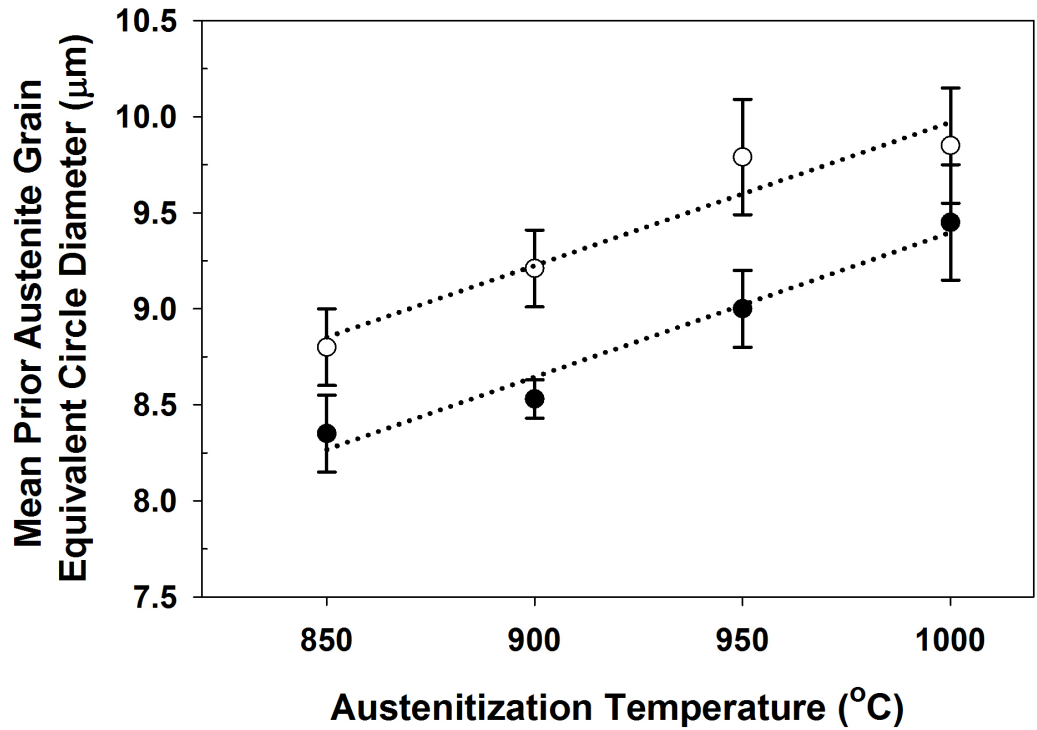

**Figure 24** The variation of mean parent austenite grain size (ECD) vs austenitization temperature after heating at 5 °C/s.

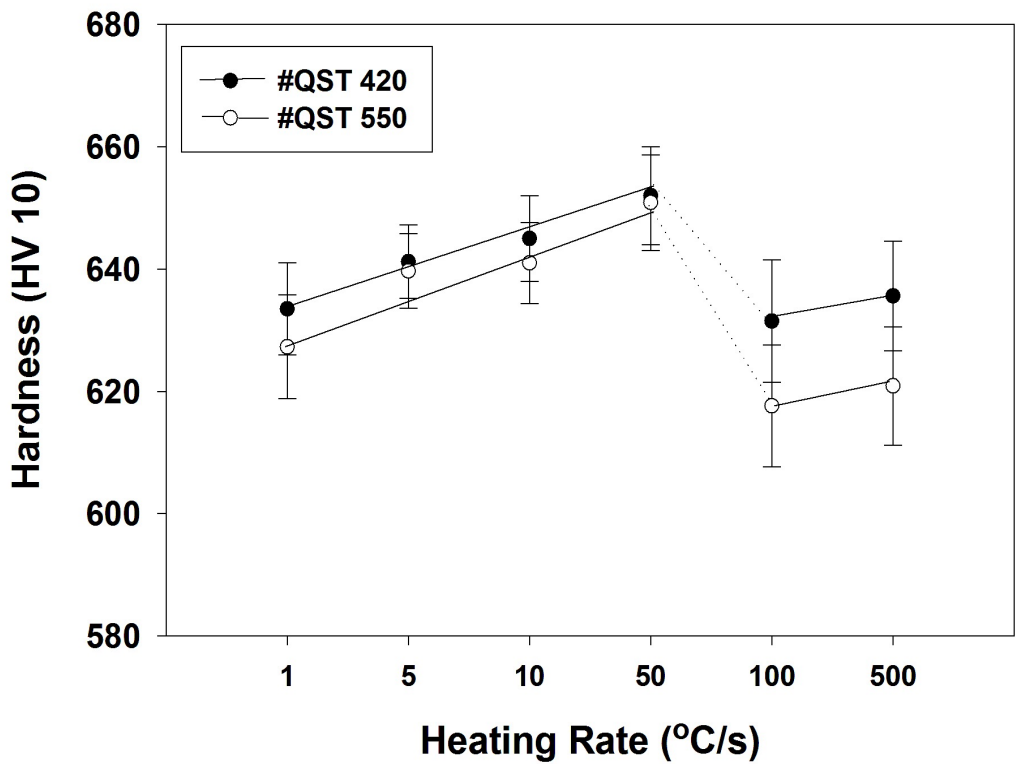

**Figure 25** The hardness variation of both samples vs heating rate. Error bars show the standard deviation for a set of 10 measurements for each sample.

3.4.2. Hardness versus Austenitization Temperature and Cooling Rate

Figure 26 gives the plots of hardness versus maximum austenitization temperature and cooling rate for both samples (heating rate of 5 °C/s). When the cooling rate is high enough to ensure an almost fully martensitic as-quenched microstructure like 40 and 60 °C/s, the variation of hardness as a function of the austenitization temperature is less than with the slower cooling rate (20 °C/s). For an almost fully martensitic microstructure, hardness slightly decreases with increasing austenitization temperature due to the coarsening of the parent austenite grains and the resultant coarsening of the martensitic structure (Figure 23 and Figure 24). But, for both initial microstructures, when the cooling rate is 20 °C/s, hardness increases significantly with increasing austenitization temperature due to the fact that, as shown below, coarser austenite has a higher hardenability leading to larger volume fractions of martensite after cooling. As shown in Figure 27 for the QST420, this is because, during cooling, lower bainite nucleates first at the austenite grain boundaries (blue arrows) and later the remaining austenite transforms to martensite. Therefore, a coarser austenite grain structure with a lower incidence of austenite grain boundaries leads to a higher fraction of martensite in the final microstructure and thereby higher hardness.

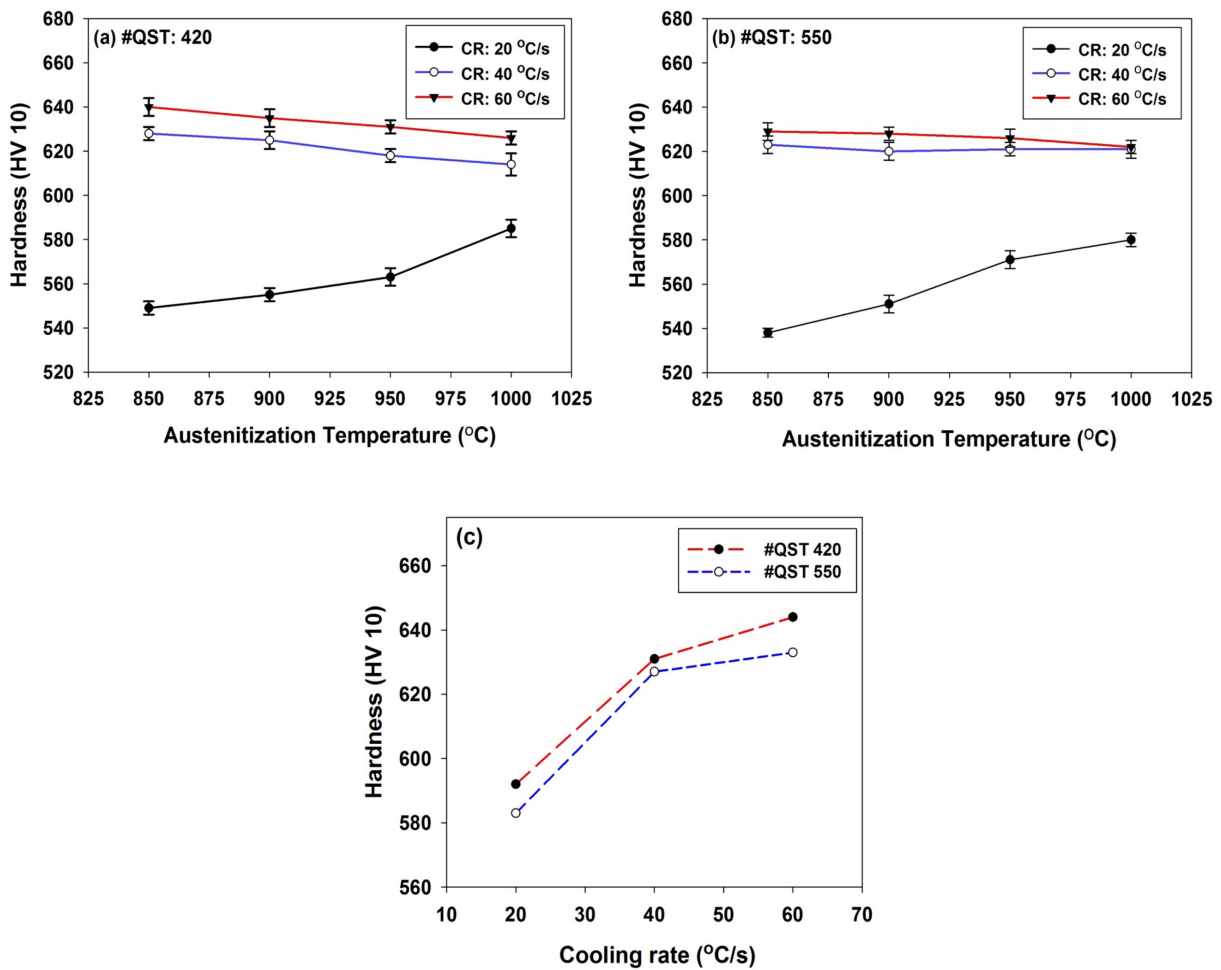

**Figure 26** Hardness versus austenitization temperature and cooling rate for a) the lower bainite starting microstructure and b) the upper lower bainite starting microstructure. c) The variation of maximum hardness as function of cooling rate for a peak temperature of 950°C.

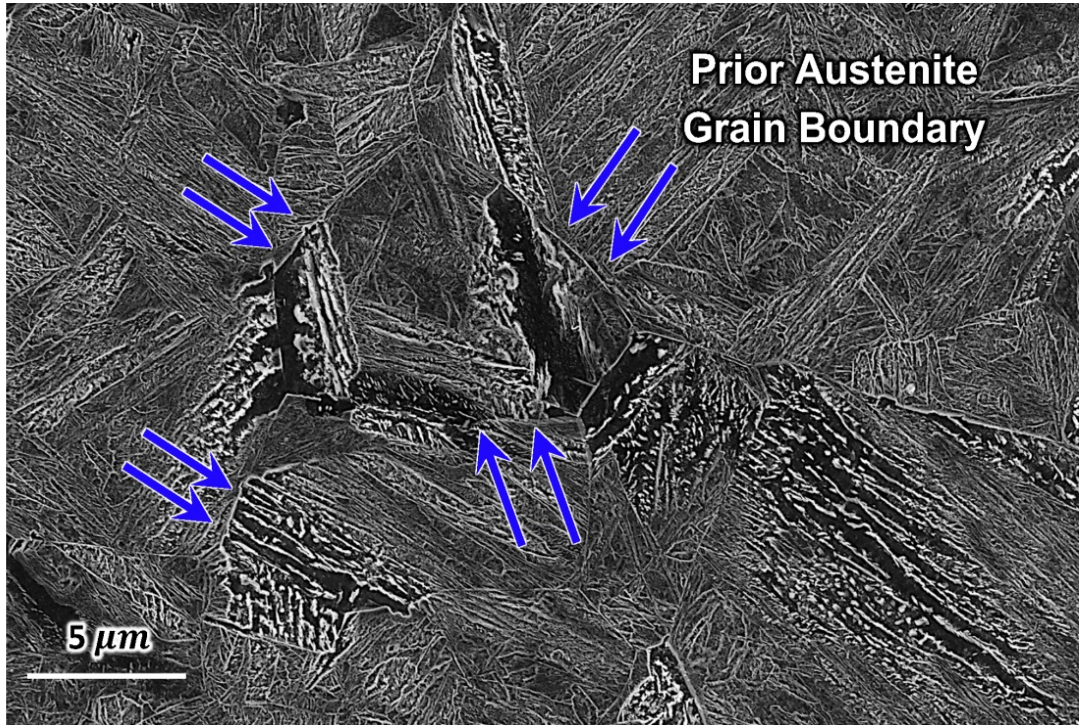

**Figure 27** FESEM micrograph of sample quenched at 20°C/s showing the formation of bainite on prior austenite grain boundaries (blue arrows).

The volume fractions of the phases present versus austenitization temperature and cooling rate for QST420 is given in Figure 28. They have been determined from the FESEM images using the point counting method with 530 points on each image and at least 3 images per case at a magnification of 5000. The results agreed with the predictions of the software JMatPro as shown in Figure 29.

*3.4.3. Hardness and Optimum Value of Induction Hardening Parameters*

From the hardness point of view, the best set of parameters for induction hardening would be re-austenitization up to a peak temperature of 850 °C at 50 °C/s and then rapid quenching at 60 °C/s. The same induction hardening process has been simulated on both samples which led to a very fine and equiaxed prior austenite grain structure as presented in Figure 30. This resulted in final hardness values of 658±5 HV and 656±6 for the QST420 and QST550 samples, respectively.

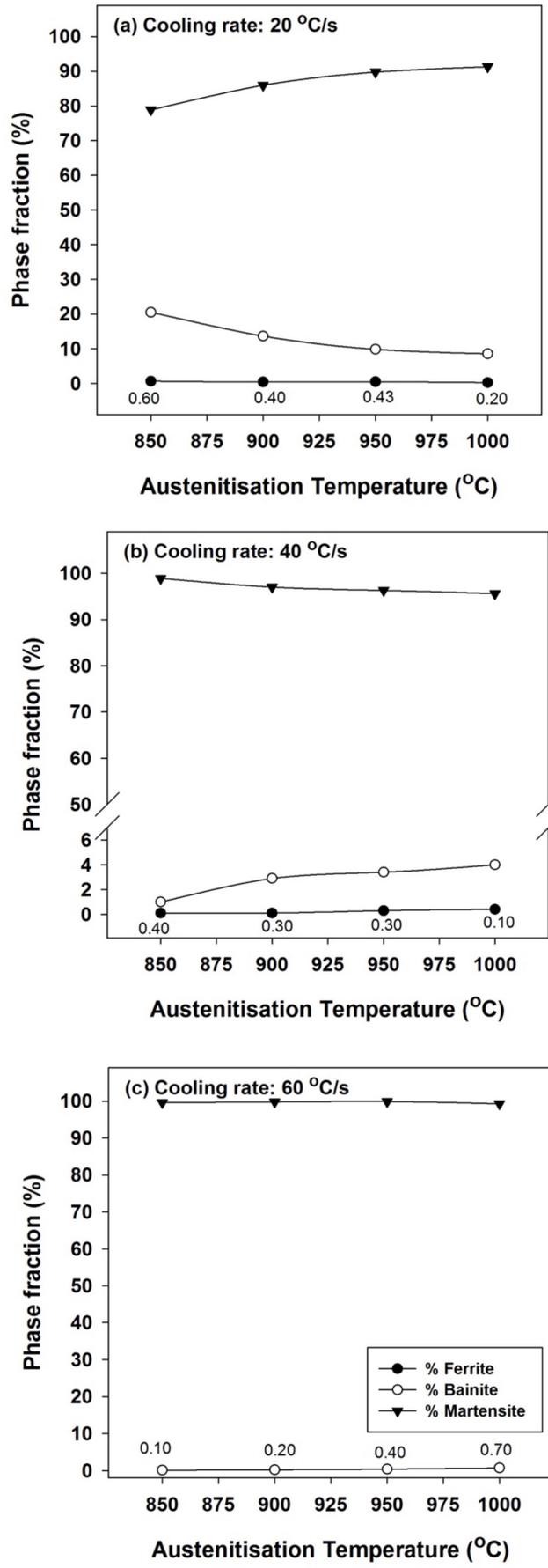

Figure 28 As-quenched phase fractions of the lower bainite sample after austenitizing at 950 °C and cooling at a) 20, b) 40 and c) 60°C/s.

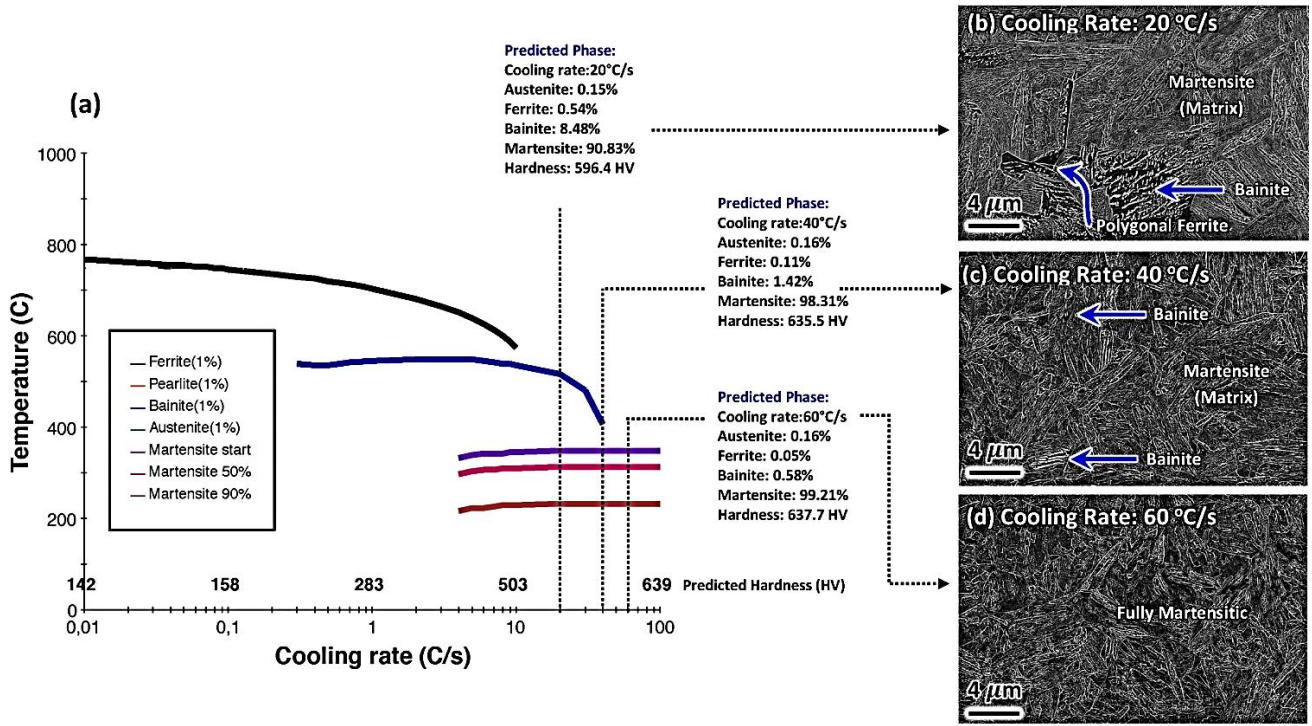

**Figure 29** a) Continuous cooling transformation (CCT) diagram along with the predicted phase fraction and hardness value for all applied cooling rates simulated using JMatPro. An example of as-quenched microstructure after austenitizing at 950 °C and cooling at b) 20, c) 40 and d) 60°C/s.

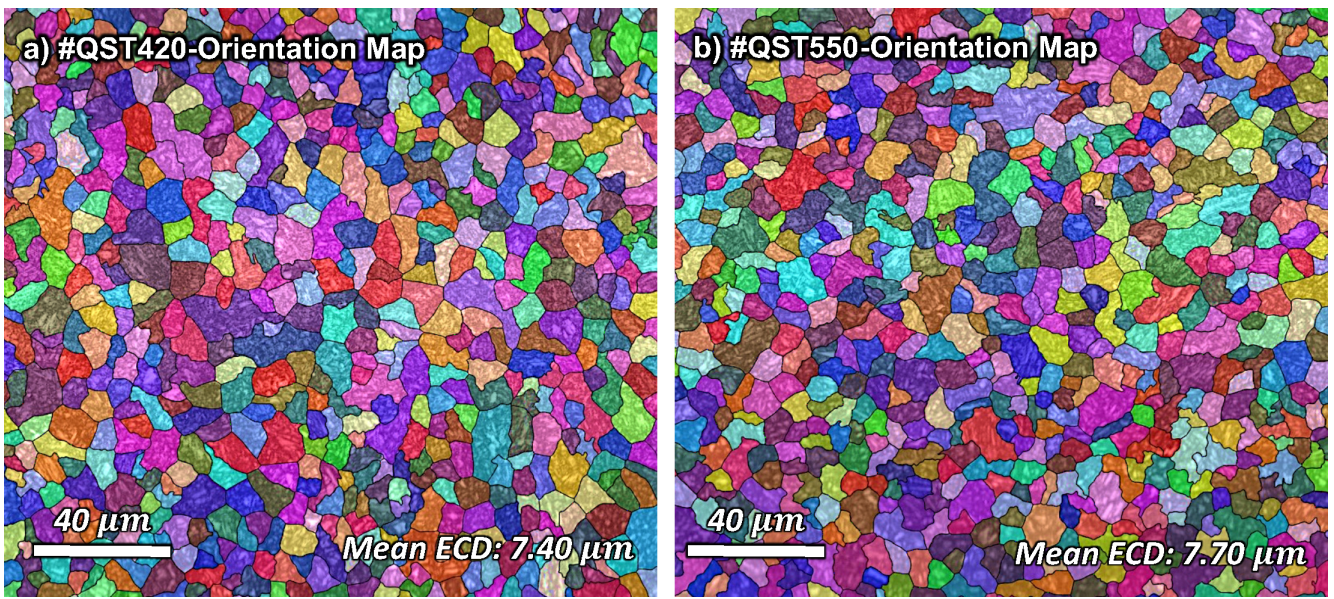

**Figure 30** The reconstructed austenite grain structure of the samples after hardening with the optimum austenitization parameters (i.e. heating rate 50 °C/s, peak temperature 850 °C and cooling rate 60 °C/s).

## 4. Conclusions

The thermodynamics and kinetics of austenitization during the simulated induction hardening of a medium carbon steel with the starting microstructures of isothermal upper and lower bainite have been studied through the use of electron microscopy, dilatometry, and confocal scanning laser microscopy. The following conclusions can be drawn from the results of these experimental methods.

- For the studied medium carbon steel with an initial bainitic microstructure, the austenite grain size achieved during the induction heating is controlled by the competition between cementite dissolution, diffusive transformation of ferrite to austenite and the initiation of the massive transformation. At very high heating rates, the massive transformation results in the appearance of coarse austenite grains in a finer grained matrix.
- An increase in the heating rate results in an increase in the start and finish temperatures of austenite formation, $A_{c1}$ and $A_{c3}$, respectively.
- During simulated induction hardening, different bainite morphologies in the starting microstructures, i.e. upper or lower bainite, only lead to slightly different final hardness levels and grain sizes.
- The influence of heating rate on $A_{c3}$ was significantly more than it was on $A_{c1}$, such that the maximum difference between $A_{c3}$ and $A_{e3}$ was almost 2.5 times more than the difference between $A_{c1}$ and $A_{e1}$.
- The austenite formation mechanism depends on the rate of heating and the initial microstructure, especially the size and chemistry of cementite.
- The highest hardness value and finest parent austenite grain size are achieved under the following condition: heating rate of 50 °C/s, austenitization peak temperature of 850 °C and immediate cooling at 60°C/s.
- The finer prior austenite structure after quenching leads to a finer martensitic microstructure provided by a sufficient cooling rate to give a fully martensitic microstructure.

**Acknowledgements**

The authors are grateful for financial support from the European Commission under grant number 675715 – MIMESIS – H2020-MSCA-ITN-2015, which is a part of the Marie Sklodowska-Curie Innovative Training Networks European Industrial Doctorate programme. The authors would also like to thank EFD induction a.s, for all the helps and support during the industrial secondment for this project especially John Inge Asperheim and Dmitry Ivanov. We would also like to express our sincere appreciation to Juha Uusitalo from University of Oulu for all his efforts and considerations with the Gleeble simulations.

**Data Availability**

The raw/processed data required to reproduce these findings cannot be shared at this time as the data also forms part of an ongoing study.